\documentstyle[multicol,aps,epsf]{revtex}

\def\twobeone{\end{multicols}
\vskip.6pc
\noindent
\vrule width3.375in height.2pt depth.2pt \vrule depth0em height1em\hfill 
\vskip.6pc}

\def\onebetwo{
\vskip.6pc
\indent
\hfill\vrule depth1em height0pt \vrule width3.375in height.2pt depth.2pt
\vskip.6pc
\begin{multicols}{2}\noindent}

\def\tup{\begin{array}{c}\uparrow\\[-1mm] -\end{array}}
\def\tdown{\begin{array}{c}\downarrow\\[-1mm]-\end{array}}
\def\bup{\begin{array}{c}-\\[-1mm]\uparrow\end{array}}
\def\bdown{\begin{array}{c}-\\[-1mm]\downarrow\end{array}}

\def\braket#1{\left|#1\right>}

\begin{document}
\draft
\title
  {One dimensional model for doubly degenerate electrons}
\author
{You-Quan Li${}^{1,2}$, Shi-Jian Gu${}^{2}$,
 Zu-Jian Ying${}^{2}$ and Ulrich Eckern${}^1$}
\address
  {${}^{1}$ Institut f\"ur Physik, Universit\"at Augsburg, 
   D-86135 Augsburg, Germany\\
   ${}^{2}$ Zhejiang Institute of Modern Physics, Zhejiang University, 
   Hangzhou 310027, China}

\date{Received: January, 2000}

\maketitle

\begin{abstract}
A Hubbard-like model with $SU(4)$ symmetry for electrons with two-fold 
orbital degeneracy is studied extensively. 
Exact solution in one dimension is derived by means of Bethe ansatz, 
where the sites are supposed to be occupied by at most
two electrons.
The features of ground state and excited states for repulsive
coupling are shown. For finite $N$ number of electrons, the configurations
of quantum numbers are given explicitly and the spectra of excitations are
obtained by solving the Bethe-ansatz equation numerically.
For infinite $N$, the ground state and various kinds of low-lying 
excitations are obtained on the basis of thermodynamics limit. 
\end{abstract}
  


\begin{multicols}{2}

\section{Introduction}

There has been much interests in the studies on
correlated electrons in the presence of orbital 
degree of freedom. 
The orbital degree of freedom is relevant to many transitional 
metal oxides \cite{McWh,Word,Bao,Pen,Feiner}.
It may be also relevant to some $C_{60}$ materials
\cite{Arovas} and samples of artificial quantum dot arrays \cite{Marston}.
For a theoretical understanding of the observed unusual properties, 
a SU(4) theory describing spin systems with orbital degeneracy was
proposed \cite{LiMSZ98,LiMSZ99}. There were also numerical \cite{Ueda} 
and perturbative \cite{Azaria} studies of 1-dimensional models for
these systems. The ground-state phase diagrams for the system with
a symmetry breaking of $SU(4)\rightarrow SU(2)\times SU(2)$ were discussed
\cite{Azaria,IQAffleck}. 
The phase separation
was recently\cite{phasesep} observed in experiment. 
Along with the rapid 
developments in experiments where the metal ions has orbital 
degeneracy in addition to spin degeneracy,
a theoretical study of such system by taking account of the kinetic terms 
caused by nearest neighbor hopping becomes important.
In a recent letter\cite{LiE99} the Hubbard model for electrons 
with orbital degeneracy was studied.
It was shown that the model not only has an 
underling SU(4) symmetry of spin-orbital double but also has
a hidden charge SU(4) symmetry.
An extended Lieb-Mattis transformation which 
maps those two SU(4) generators into each other is given. 
On the basis of elementary degenerate
perturbative theory,
it was shown that the effective Hamiltonian with strong
repulsive coupling at half-filling is equivalent to 
the Hamiltonian of the $SO(6)$ 
Heisenberg model, and that at quarter-filling is equivalent to the one 
of SU(4) Heisenberg model.
Some features of the model in one dimension was also briefly described.

In present paper, we study the one-dimensional Hubbard-like model 
with SU(4) symmetry for electrons with two-fold orbital 
degeneracy extensively. 
Its exact solution is formulated by means of Bethe ansatz 
if sites are assumed to be occupied by at most two electrons.
The features of the ground state and excited states for repulsive
coupling are shown. For finite $N$ number of electrons, the configurations
of quantum numbers are given explicitly and the spectra of excitations are
obtained by solving the Bethe-ansatz equation numerically.
For infinite $N$, the ground state and various kind of low-lying 
excitations are obtained on the basis of thermodynamics limit. 

The paper is organized as follows. 
In next section we propose the model 
Hamiltonian with some interpretative remarks. Employing standard method
we carefully formulated the first quantized version of the Hamiltonian.
We make an allowed modification so that the Bethe-Yang ansatz 
\cite{Yang67,Sutherland68} is applicable to this model. 
A detailed formulation from the Bethe-ansatz wave function to the 
Bethe-ansatz equation is given.
In Sec. \ref{sec:ground}, we explicitly show how the quantum numbers
in the Bethe-ansatz equation should be taken for the non-degenerated ground
state. We calculate the ground-state energy and Fermi momentum numerically
for different numbers of electrons. We also compare them for different 
coupling constants.
In Sec. \ref{sec:excite}, we study the excited states extensively by 
analyzing the possible variations in the sequence of quantum numbers.
We indicate in each case how the quantum numbers change from integers
to half-integers (or vise versa) with respect to that of the ground state.
Numerical results of energy-momentum spectra for each excitation
are given there.
In Sec. \ref{sec:special}, two special cases, weak and strong coupling
are discussed. We are able to deduce several interesting properties from 
the Bethe-ansatz equation without solving it directly.
In Sec. \ref{sec:thermodynamics}, we consider the thermodynamics limit. 
After giving some general formulae and expressions, we study the ground state
and calculate the ground-state energy explicitly for strong coupling.
In Sec. \ref{sec:spin-orbital}, we discuss low-lying excitations in the
spin-orbital sector on the basis of thermodynamics limit. 
Both contributions of holes and 2-strings\cite{Woynarovich}
are taken into account. The singlet
excitation and several multiplet excitations are obtained.
In Sec. \ref{sec:charge}, we discuss low-lying excitations in charge 
sector by thermodynamics limit. The holon-antiholon and holon-holon
excitations are obtained.
Sec. \ref{sec:conclusion} is a summary of the main results of the 
paper and some conclusive discussions.

\section{The model and its exact solution}\label{sec:exact}

We consider a Hubbard-like model for electrons with two-fold orbital degeneracy.
The spin components are denoted by up $(\uparrow)$ and down $(\downarrow)$,
the orbital components by top and bottom. The four possible states of
electrons are
\begin{eqnarray} 
|1>=|\tup >, \;\;\;&  
|2>=|\tdown >, \nonumber\\[0mm] 
|3>=|\bup >, \;\;\;& 
|4>=|\bdown >.
\end{eqnarray}
We use $1, 2, 3,$ and $4$ to label these states from now on. 
The model Hamiltonian then reads
\begin{equation}
H=-t\sum_{i,a}(C^+_{i,a}C_{i+1,a}+C^+_{i+1,a}C_{i,a})
   +U\sum_{i,a<a'}n_{i,a}n_{i,a'} 
\label{eq:Hamiltonian}
\end{equation}
where $i=1,2,...,L$ identify the lattice site, $a, a'= 1, 2, 3, 4$ 
specify the spin and orbital as defined in the above.
The $C^+_{i a}$ creates an electron  
with spin-orbital component $a$ on site $i$, and 
$n_{i a}:=C^{+}_{i a}C_{i a}$ is the corresponding number operator
at site $i$. 
Eq.(\ref{eq:Hamiltonian}) is the Hamiltonian for four-component systems,
and there were various discussions on multi-component Hubbard
model in one dimension \cite{Schlottmann,Choy,Assaraf,MAffleck,FrahmS}.
Whereas the physics that
eq.(\ref{eq:Hamiltonian}) describes will be precise only when the 
representation space for the internal degree of 
freedom is specified \cite{LiE99}.
It is specified to the spin and orbital, and the site is assumed to
be occupied by at most two electrons in our present model.

It is convenient to consider
the states that span the Hilbert space of $N$-particles
\[
  |\psi > =\sum_{\{a_j\},\{x_j\}}
       \psi_{a_1,\cdots a_N }(x_1,\cdots, x_N)    
          C^{+}_{x_1 a_1}\cdots C^{+}_{x_N a_N}|0>.
\]
where $x_j\in\{1,2,...,L\}$, $a_j\in\{1,2,3,4\}$ and $j=1,2,...,N$. 
The eigenvalue problem 
$H |\psi> = E |\psi>$
becomes an $N$-particle quantum mechanical problem
with the first quantized Schr\"odinger operator (Hamiltonian),
\begin{equation}
{\cal H} = -t\sum_{j=1}^{N}\Delta_j  
       +U\sum_{i<j}(1-\delta_{a_i,a_j})\delta(x_i,x_j),    
\label{eq:Schoperator}
\end{equation}
if site occupations of more than two electrons are excluded 
\cite{ChoyHaldane}, where 
$\Delta_j\psi:=\psi(\cdots,x_j+1,\cdots)
+\psi(\cdots,x_j-1,\cdots)$.

The wave function of Bethe-ansatz form in the region  
$x\in{\cal C}(Q):=\{x|\, 1\leq x_{Q1}<\cdots<x_{QN}\leq L\}$
reads
\begin{equation}
\psi_a(x)=\sum_{P\in S_N}A_a(P,\,Q)e^{i(Pk|x)},
\label{eq:BAW}
\end{equation}
where $x=(x_1,x_2,...,x_N)$, 
$a:=(a_{1}$, $a_{2},\cdots,a_{N})$, 
$a_j$ stands for the spin-orbital component of the ``$j$th'' particle; 
$Pk$  is the image of a given 
$k:=(k_1, k_2,\cdots, k_N )$ by 
a mapping $P\in S_N$;
$S_N$ denotes the permutation group of $N$ objects; 
$(Pk|x)=\sum_{j=1}^{N} (Pk)_j (x)_j$. 
The coefficients $A(P,Q)$ 
are functionals on $S_N\otimes S_N$.
We known that any permutations can always be 
expressed as the product of the neighboring interchanges 
$\Pi^j: (\cdots,z_j, z_{j+1},\cdots)\mapsto(\cdots,z_{j+1},z_j,\cdots)$.
So the requirement of antisymmetry for fermionic permutation is
$(\Pi^j\psi)_a(x)= -\psi_a(x)$,
which gives  
\begin{equation}
A(P,\,\Pi^jQ)= 
-{\cal P}^{Qj,Q(j+1)}A(\Pi^jP,\,Q),
\label{eq:antisymmetry}
\end{equation}
where the spin-orbital labels are omitted and
${\cal P}^{Qj,Q(j+1)}$ is the SU(4) spinor representation 
of the permutation $\Pi^j$ operator.
An immediate consequence of (\ref{eq:antisymmetry}) is 
$\delta_{a_i,a_j}\delta(x_i,x_j)\psi_a(x)=0$, and hence
we are allowed to consider, instead of (\ref{eq:Schoperator}), the 
following equivalent Schr\"odinger operator
\begin{equation}
{\cal H} = -t\sum_{j=1}^{N}\Delta_j  
       +U\sum_{i<j}\delta(x_i,x_j),   
\label{eq:theSchoperator}
\end{equation}
here the interaction terms are independent of the spin-orbital labels.
Then the strategy of ref.\cite{Yang67,LiebWu} can be used to solve the 
wave functions. The S-matrix that relates 
the coefficients $A's$ between distinct regions in the 
configuration space of $N$ electrons can be solved from the Schr\"odinger
equation in the vicinity of hyperplane with $(Qx)_j=(Qx)_{j+1}$.
Accordingly, we get
\begin{equation}
S^{Qj,Q(j+1)}=
  \frac{\sin(Pk)_{Qj}-\sin(Pk)_{Q(j+1)}+ic{\cal P}^{Qj,Q(j+1)}}
        {\sin(Pk)_{Qj}-\sin(Pk)_{Q(j+1)}+ic}, 
\label{eq:Smatrix}
\end{equation}
where $2c=U/t$. As it satisfies Yang-Baxter equation \cite{Yang67}, 
the Bethe-ansatz wave function is then consistently determined, 
i.e., the coefficients $A$'s in any region are determined 
up to an overall factor by the 
$\check{S}^{Qj,Q(j+1)}:=-{\cal P}^{Qj,Q(j+1)}S^{Qj,Q(j+1)}$
and that in different regions are related by (\ref{eq:antisymmetry}). 
If let $c\rightarrow\infty$ in (\ref{eq:Smatrix}), we know the wave function
will be null if there are two $k$'s being the same value. So in the 
strong coupling limit, the $k$'s must take distinct values though there 
can be four states corresponding to the same $k_j$ in the absence of
interaction.

The periodic boundary condition is guaranteed provided that 
$A(P,\gamma Q)e^{i(Pk)_{1}L }=A(P,Q)$ in which
$\gamma = \Pi^{N-1}\Pi^{N-2}\cdots\Pi^{2}\Pi^{1}$.
After applying  the S-matrices successively, one obtain an eigenvalue 
equation in the SU(4) spinor space:
\begin{eqnarray}
\,&\,&S^{Q1,QN}S^{Q1,Q(N-1)}\cdots S^{Q1,Q2}A(P, Q)\nonumber\\
\,&\,& =e^{-i(Pk)_1L}A(P, Q).
\label{eq:PBC}
\end{eqnarray}
The eigenvalue problem (\ref{eq:PBC}) can be diagonalized by means of 
quantum inverse scattering method\cite{Faddeev} by defining the 
transfer matrix as 
$t(\alpha)=tr{\cal T}(\alpha)$ where
${\cal T}(\alpha)=T_{AN}(\alpha-\alpha_N)\cdots
   T_{A2}(\alpha-\alpha_2) T_{A1}(\alpha-\alpha_1)$,
$T_{Aj}(\alpha):=S^{Aj}(\alpha)\in End(V^A\otimes V^j)$.
$End$ means endomorphism.
It can also be diagonalized by similar procedure as in 
ref.\cite{Sutherland68} where the general case of 
continuous model with $\delta$-function interaction was solved.
The Bethe-ansatz equations reads
\twobeone
\begin{eqnarray}
\,&\,& e^{ik_jL}
     =\prod_{\alpha=1}^{M}\Xi_{-1/2}(\sin k_j-\lambda_\alpha),
       \nonumber\\
1&=&-\prod_{l=1}^{N}\Xi_{-1/2}(\lambda_\alpha-\sin k_l)
      \prod_{\alpha'=1}^{M}\Xi_1(\lambda_\alpha-\lambda_{\alpha'})
       \prod_{\beta=1}^{M'}\Xi_{-1/2}(\lambda_\alpha-\mu_\beta),
         \nonumber\\
1&=&-\prod_{\alpha=1}^{M}\Xi_{-1/2}(\mu_\beta-\lambda_\alpha)
    \prod_{\beta'=1}^{M'}\Xi_1(\mu_\beta-\mu_{\beta'})
      \prod_{\gamma=1}^{M''}\Xi_{-1/2}(\mu_\beta-\nu_\gamma),
       \nonumber\\
1&=&-\prod_{\beta=1}^{M'}\Xi_{-1/2}(\nu_\gamma-\mu_\beta)
      \prod_{\gamma'=1}^{M''}\Xi_1(\nu_\gamma-\nu_{\gamma'}),
\label{eq:BAE}
\end{eqnarray}
\onebetwo
where $\Xi_n(x):=[x+inc]/[x-inc]$.
Eq.(\ref{eq:BAE}) was given firstly in \cite{Choy}. 
Because a particular chemical potential was introduced in the Hamiltonian,
the Bethe-ansatz equation derived in \cite{FWadati} is the same as the
SU(4) Heisenberg model \cite{LiMSZ99}.
We write out the Bethe-ansatz equation in a form so that 
it is easy to be remembered by means of the ``Dynken diagram''\cite{Gilmore}
of $A_3$ Lie algebra
\begin{figure}
\setlength{\unitlength}{1mm}
\begin{picture}(66,18)(-6,0)
\linethickness{0.5pt}
\put(2,6){\circle*{2}}\put(1,8){$k$}\put(0.5,2){${\small N}$}
  \put(3,6){\dashbox{1.0}(8,0)}
\put(12,6){\circle{2}}\put(11,8){$\lambda$}\put(10,2){${\small M}$}
  \put(13,6){\line(1,0){7.8}}
\put(22,6){\circle{2}}\put(21,8){$\mu$}\put(20,2){${\small M'}$}
  \put(23,6){\line(1,0){7.8}}
\put(32,6){\circle{2}}\put(31,8){$\nu$}\put(30,2){${\small M''}$}
\end{picture}
\end{figure}
\noindent
Where the dark dot is added to represent the charge rapidity $k_j$
which also has an angle of  $120^o$ relative to the first simple root
$r_1$.
The subscripts in $\Xi$ in eq.(\ref{eq:BAE}) are then related to the 
covariant components of the simple roots when the simple roots are 
chosen as non-orthogonal basis, accordingly,
$r_1=(-1/2,\, 1,\, -1/2,0)$, 
$r_2=(0,\, -1/2,\, 1, -1/2)$,
$r_3=(0, 0, -1/2,\, 1)$.
The highest weight vector ${\bf w}=(w_1, w_2, w_3)$ labeling
the representation of SU(4) carried out by the corresponding eigenstates is 
given by 
\begin{eqnarray}
w_1 &=& N-2M+M',\nonumber\\
w_2 &=& M-2M'+M'',\nonumber\\
w_3 &=& M'-2M''.
\label{eq:HWvector}
\end{eqnarray}

A set of coupled transcendental equations
are derived by taking the logarithm of (\ref{eq:BAE}),
\twobeone
\begin{eqnarray}
k_j-\frac{1}{L}\sum_{\alpha=1}^{M}\Theta_{-1/2}(\sin k_j-\lambda_\alpha)
     &=&\frac{2\pi}{L}h_j,
      \nonumber\\
\sum_{l=1}^{N}\Theta_{-1/2}(\lambda_\alpha-\sin k_l)
  +\sum_{\alpha'=1}^{M}\Theta_1(\lambda_\alpha-\lambda_{\alpha'})
   +\sum_{\beta=1}^{M'}\Theta_{-1/2}(\lambda_\alpha-\mu_\beta)
     &=&-2\pi I_\alpha,     
      \nonumber\\
\sum_{\alpha=1}^{M}\Theta_{-1/2}(\mu_\beta-\lambda_\alpha)
  +\sum_{\beta'=1}^{M'}\Theta_1(\mu_\beta-\mu_{\beta'})
    +\sum_{\gamma=1}^{M''}\Theta_{-1/2}(\mu_\beta-\nu_\gamma)
      &=&-2\pi J_\beta,
       \nonumber\\
\sum_{\beta=1}^{M'}\Theta_{-1/2}(\nu_\gamma-\mu_\beta)
  +\sum_{\gamma'=1}^{M''}\Theta_1(\nu_\gamma-\nu_{\gamma'})
    &=&-2\pi K_\gamma,
\label{eq:logBAE}
\end{eqnarray}
\onebetwo\noindent
where $\Theta_n(x):=2\tan^{-1}(\frac{x}{nc})$. The quantum number 
$h_j$ for charge rapidity $k_j$ takes integer or half-integer value
depending on whether $M-1$ is odd or even. The quantum numbers 
$I_\alpha$, $J_\beta$ or $K_\gamma$ for flavor 
(we refer for the spin-orbital double) rapidities
$\lambda_\alpha$, $\mu_\beta$ or $\nu_\gamma$,
take integer or half-integer values respectively depending on 
whether $N-M-M'$, $M-M'-M''$ or $M'-M''$ is odd or even.
These properties arise from the logarithm function. Once the roots are
solved from the above equation (\ref{eq:logBAE}), the energy and momentum
will be calculated by
\begin{eqnarray}
E&=&-2t\sum_{j=1}^N\cos k_j, \nonumber\\ 
P&=&\frac{2\pi}{L}
  \left[\sum^N_{l=1}h_l+\sum^M_{\alpha=1}I_\alpha
   +\sum^{M'}_{\beta=1}J_\beta+\sum^{M''}_{\gamma=1}K_\gamma\right].
\label{eq:EandPsum}
\end{eqnarray}

\section{Ground state}\label{sec:ground}

\begin{figure}
\epsfclipoff
\fboxsep=0pt
\setlength{\unitlength}{1mm}
\begin{picture}(150,180)(0,0)
\linethickness{1pt}
\epsfysize=8cm
\put(18,88){{\epsffile{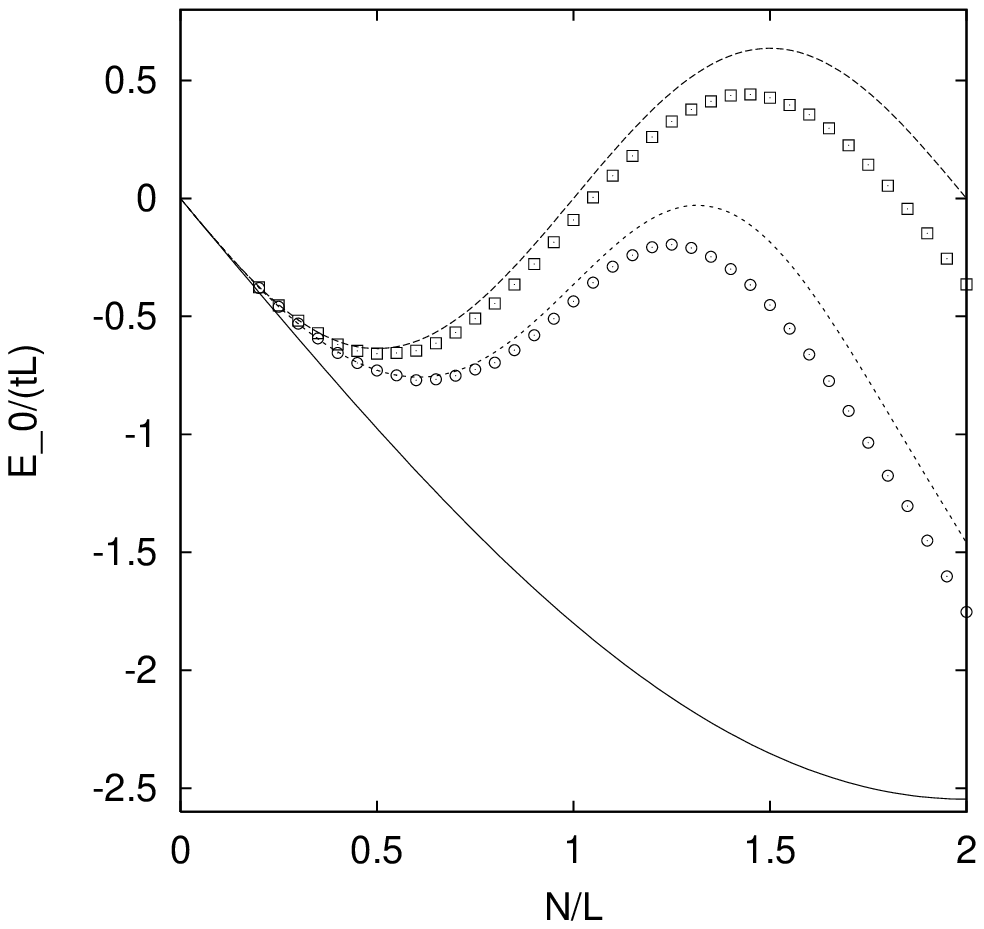}}}
\epsfysize=8cm
\put(18,1){{\epsffile{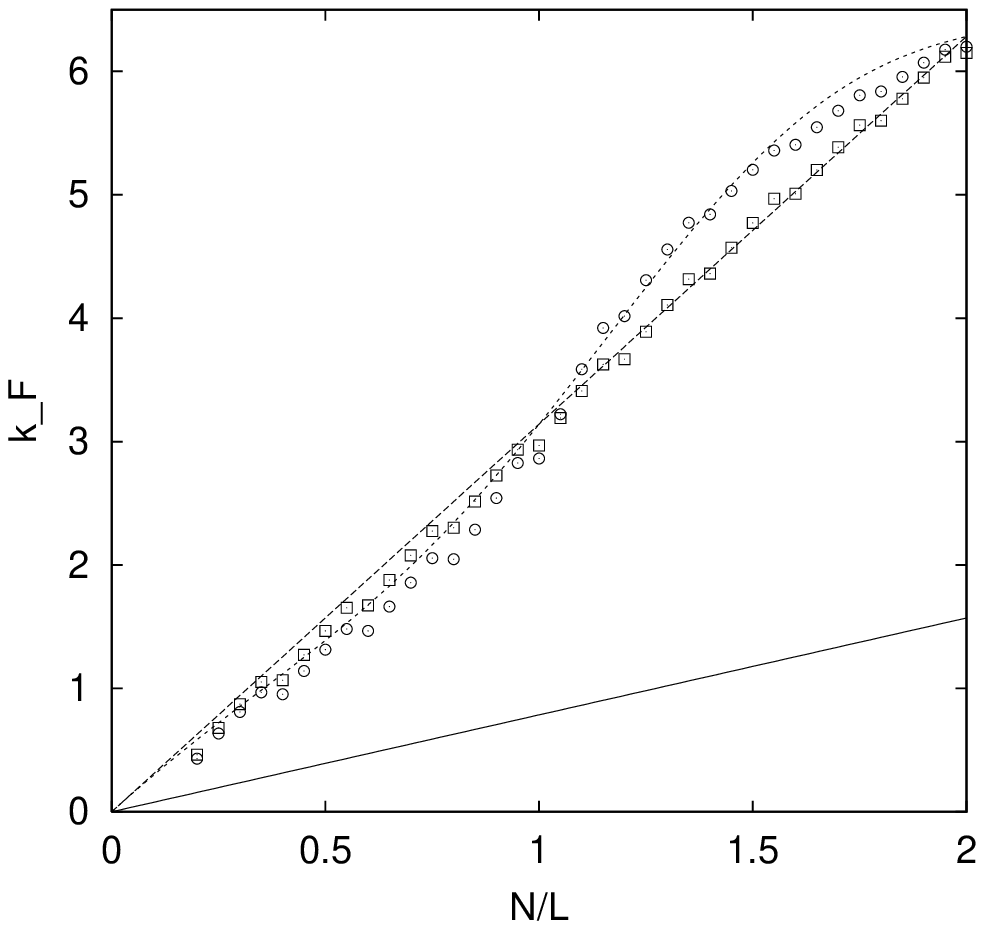}}}
\end{picture}
\vspace{12mm}
\caption{The relations between filling factor $N/L$ 
and ground-state energy $E_0$
(upper figure) or the Fermi momentum $k_F$ (lower figure). 
The points are calculated with $U/t=10$ ($\Box$) and $U/t=2$ ({\large$\circ$})
by taking $N$ from 4 to 40 for $L=20$.
Noninteracting case is plotted by solid lines. The other
lines (dashed lines for $U/t=\infty$ and dotted lines for $U/t=10$) 
are plotted from the results of thermodynamics limit.
}
\label{fig:ground}
\end{figure}

The ground state is nondegenerate only if $N=4n$ for $n$ being odd numbers.
This is easily seen by considering non-interaction case. The momentum
eigenvalues (with periodic boundary condition) of non-interacting electrons
are $k=m(2\pi/L)$, $m=0, \pm 1, \pm 2 \cdots$. For example, $N=4n$ for
$n=even$, the ground state has a 70-fold degeneracy.  In the following we 
will restrict ourselves to the case of $N_0=4n$ for $n=odd$, and consider the 
non-degenerated ground state. The non-degenerated ground state is a SU(4)
singlet which is characterized by a 4-row and n-column Young tableau.
The quantum numbers in (\ref{eq:logBAE}) for the ground state are
\begin{eqnarray}
\{h_j\}&=&\{-(N_0-1)/2, ..., (N_0-1)/2\}, \nonumber\\
\{I_\alpha\}&=&\{-(3N_0-4)/8, ..., (3N_0-4)/8\}, \nonumber\\
\{J_\beta\}&=&\{-(N_0-2)/4, ..., (N_0-2)/4\}, \nonumber\\
\{K_\gamma\}&=&\{-(N_0-4)/8, ..., (N_0-4)/8\}.
\label{eq:QNground}
\end{eqnarray}
Obviously, $h$'s and $J$'s are consecutive half-integers 
while $I$'s and $K$'s are consecutive integers. 
As a result of (\ref{eq:QNground}) and (\ref{eq:EandPsum}), 
the momentum of the non-degenerate
ground state is zero.
We plot the ground-state energy with respect to the filling factor 
$N/L$ for various coupling constant in fig.(\ref{fig:ground}). 
The relation between Fermi momentum and
the filling factor for different coupling constant is also plotted there.
These numerical results are calculated for $L=20$ and $N$ from $4$ to $40$.
For $N\rightarrow\infty$, we can take thermodynamics limit
which will be discussed in Sec.\ref{sec:thermodynamics}.
The ground state for $N=N_0+1$, $N=N_0+2$ etc. are degenerate about which
we will discuss later.

\section{Excited states}\label{sec:excite}

\subsection{Excitations above the nondegenerate ground state}

\begin{figure}
\epsfclipoff
\fboxsep=0pt
\setlength{\unitlength}{1mm}
\begin{picture}(150,180)(0,0)
\linethickness{1pt}
\epsfysize=8cm
\put(18,88){{\epsffile{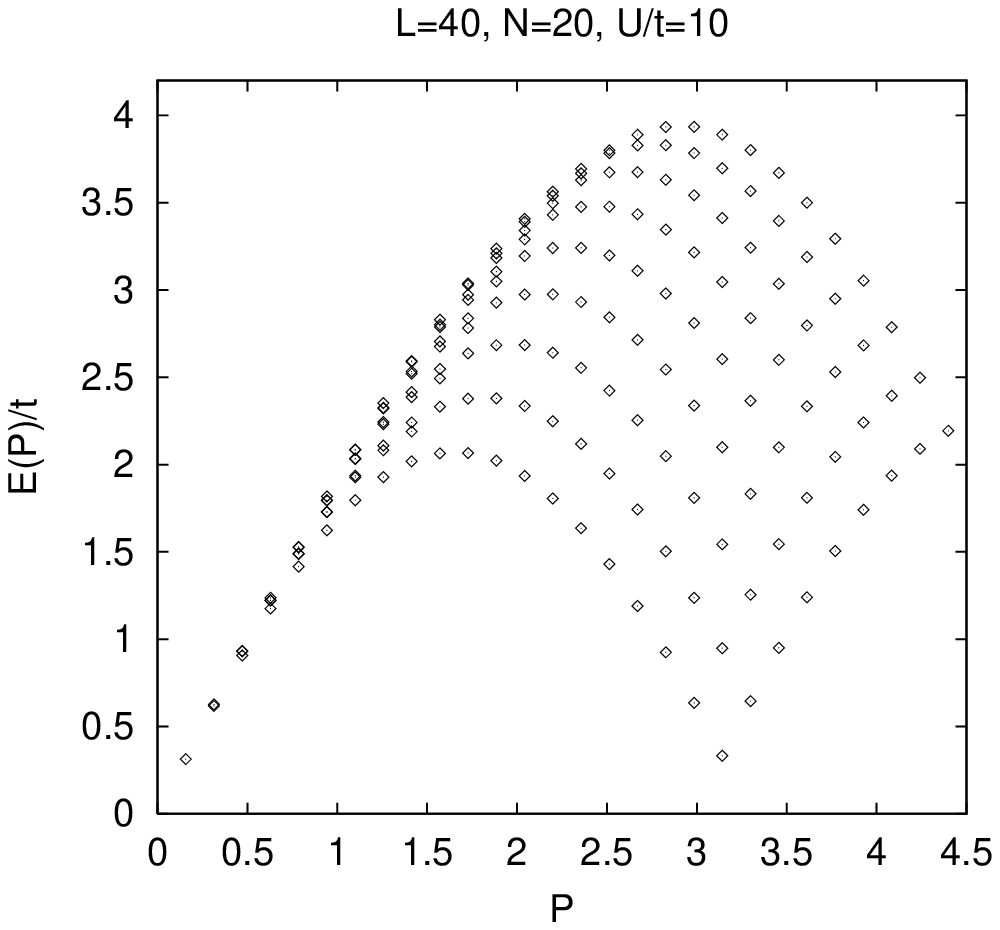}}}
\epsfysize=8cm
\put(18,1){{\epsffile{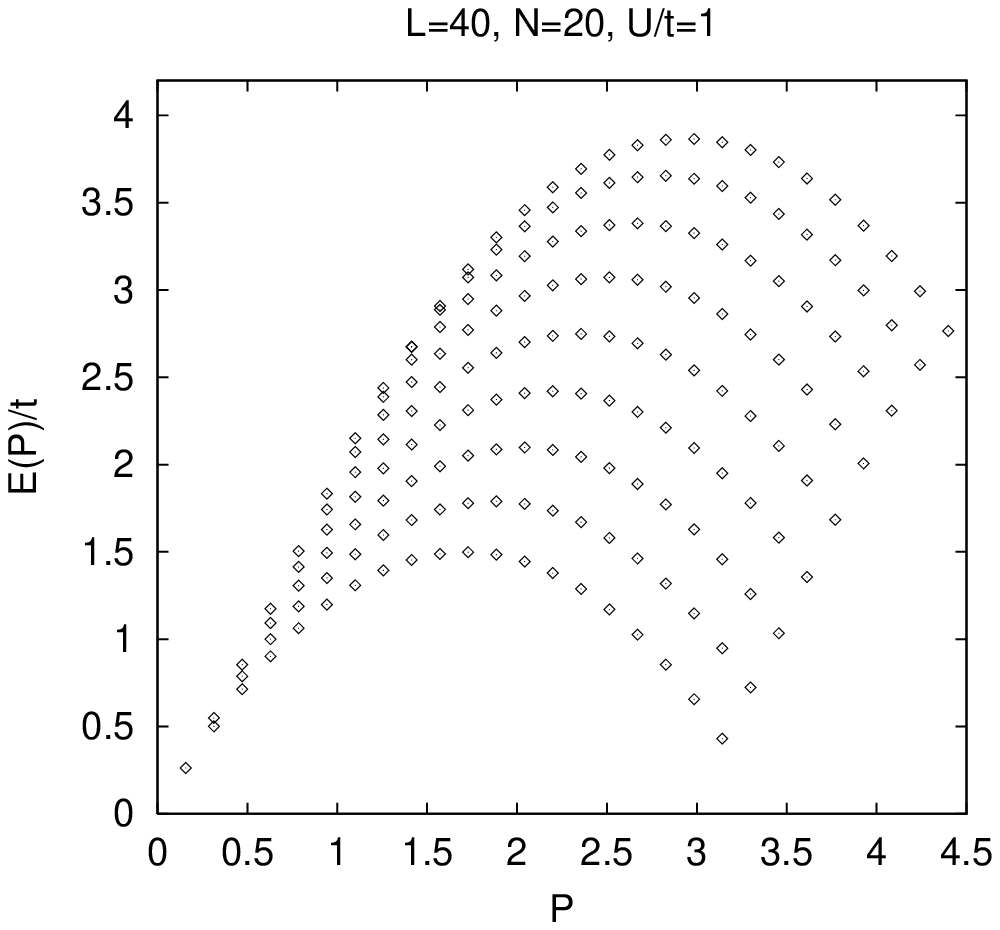}}}
\end{picture}
\vspace{12mm}
\caption{``particle''-hole excitation spectrum for the model
of 40 sites with 20 electrons. The overall structure of the spectrum
does not change much between the strong ($U/t=10$) and weak ($U/t=1$)
correlation strengths. }
\label{fig:ph}
\end{figure}

The excited states are obtained by variation in the sequence of 
quantum numbers $\{h_j\}$, $\{I_\alpha\}$, $\{J_\beta\}$ or $\{K_\gamma\}$
from that for the ground state. The simplest case is to remove one of the 
$h$'s from the sequence of ground state (\ref{eq:QNground}) and add a new
$h_0$ outside of the original sequence. That is 
\begin{eqnarray}
\{h_j\}=\{-\frac{N_0 -1}{2},...\frac{N_0 -1}{2}+n_0-1, \nonumber\\
   \frac{N_0 -1}{2}+n_0+1,...,\frac{N_0 -1}{2}, h_0\},
\label{eq:QNph}
\end{eqnarray} 
with $|h_0|>(N_0-1)/2$ and the other sequences in (\ref{eq:QNground}) 
keep unchanged. Clearly, the $(N_0-1)/2+n_0$ is absent in the set 
(\ref{eq:QNph}). 
In fig.(\ref{fig:ph}), we plot the numerical results for energy-momentum 
spectrum, which is a two-parameter family of excitation.
This kind of excitations are singlet states.

\begin{figure}
\epsfclipoff
\fboxsep=0pt
\setlength{\unitlength}{1mm}
\begin{picture}(150,180)(0,0)
\linethickness{1pt}
\epsfysize=8cm 
\put(18,88){{\epsffile{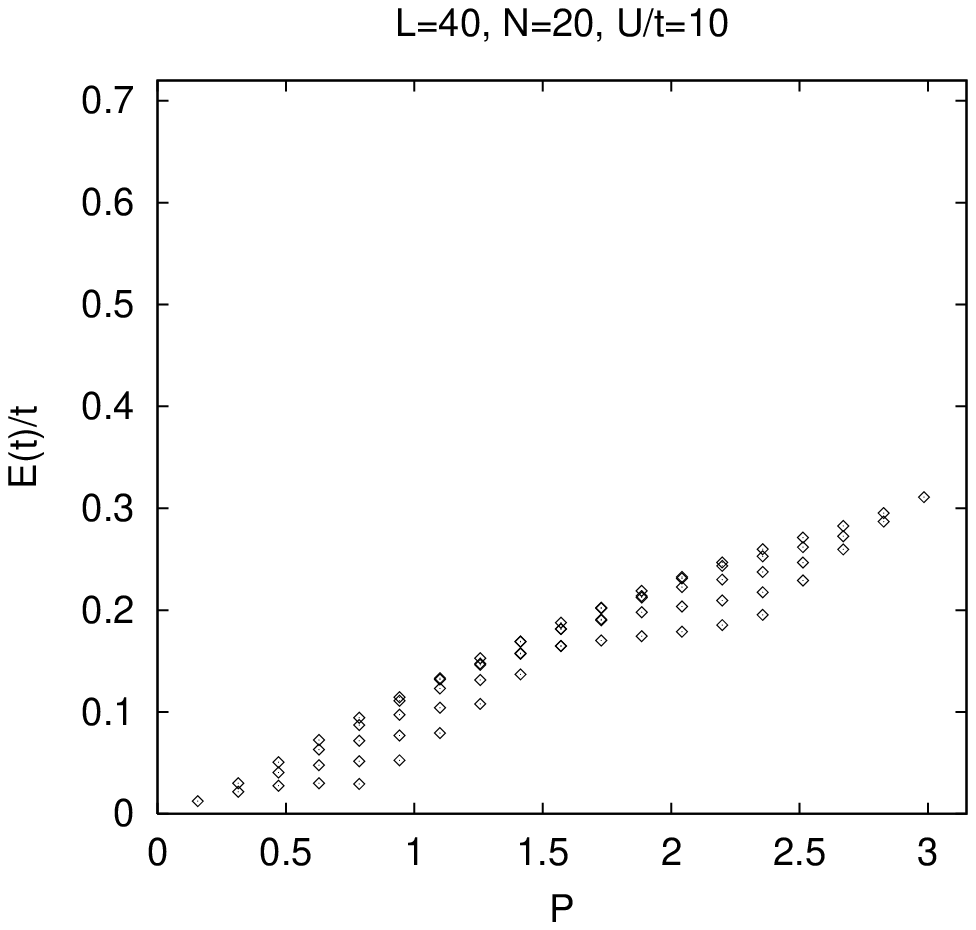}}}
\epsfysize=8cm
\put(18,1){{\epsffile{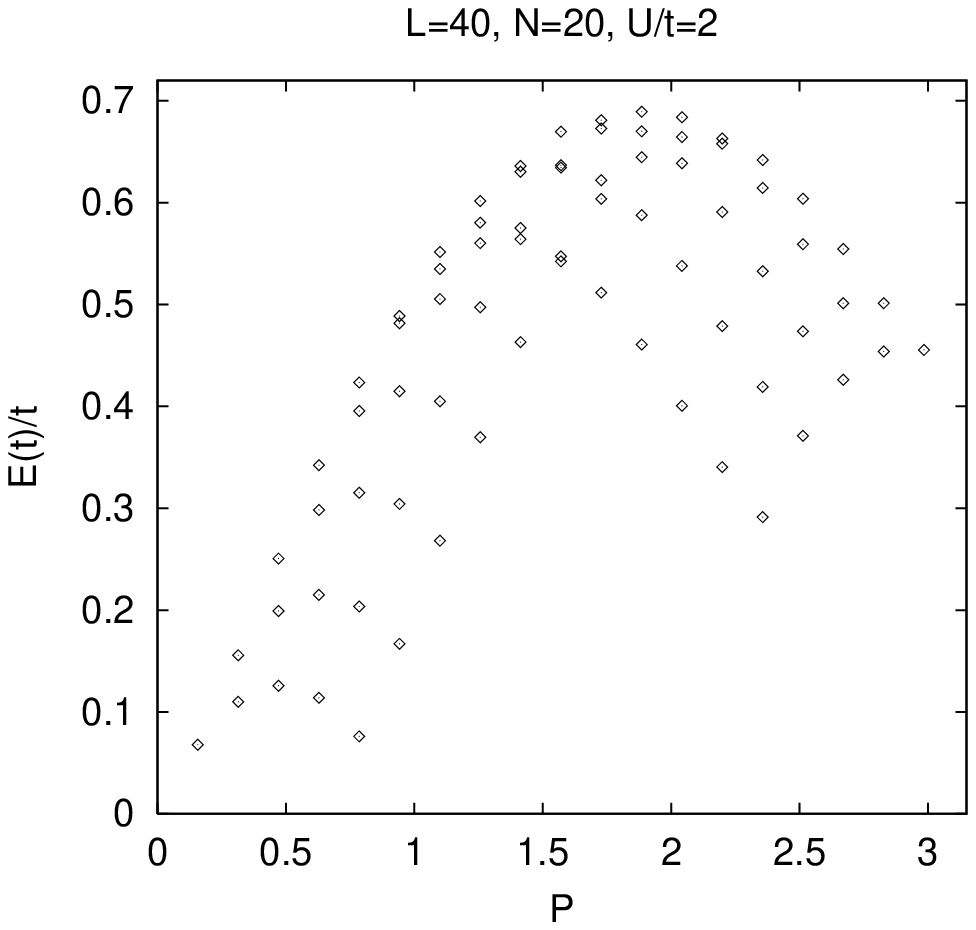}}}
\end{picture}
\vspace{12mm}
\caption{15-fold excitation spectrum for the model
of 40 sites with 20 electrons. The overall structure of the spectrum
changes much according to the correlation strengths.
The lowest excitation energy and the whole pattern
are raised when the correlation strength
decreased.}
\label{fig:so15}
\end{figure}

There are several further possibilities. 
After moving one box from the fourth
row in the Young tableau of the ground state to the first row,
we get a Young tableau labeling a 15-dimensional 
irreducible representation of 
SU(4) according to the knowledge of group theory. It requires 
$M=3N_0/4-1$, $M'=N_0/2-1$ and $M''=N_0/4-1$. This causes the 
$h$'s and $J$'s to take integer values instead of half-integer values that
were taken for ground state. There are now $M$ allowed values for the 
$M-1$ distinct $I$'s, and $M''$ allowed values for the $M''-1$
distinct $K$'s. Consequently, holes in the $I$'s and $K$'s 
sequences indispensably occurred, and then the low-energy states are
parameterized by,
\begin{eqnarray}
\{h_j\}&=&\{-N_0/2+1, ..., N_0/2-1, N_0/2\},\nonumber\\
\{J_\beta\}&=&\{-N_0/4+1,...,N_0/4-1\},\nonumber\\
I_1&=&-\frac{3N_0-4}{8}+\delta_{1,\alpha_1},\nonumber\\
I_\alpha &=&I_{\alpha-1}+1+\delta_{\alpha,\alpha_1}\,\,
 (\alpha=2,...\frac{3N_0}{4}-1),\nonumber\\
K_1 &=&-\frac{N_0-4}{8}+\delta_{1,\gamma_1}, \nonumber\\
K_\gamma &=& K_{\gamma-1}+1+\delta_{\gamma,\gamma_1}\,\,
 (\gamma=2,...,\frac{N_0}{4}-1),
\label{eq:QN15}
\end{eqnarray}
where $1\leq\alpha_1\leq 3N_0/4$, $1\leq\gamma_1\leq N_0/4$. 
Numerical results of the energy-momentum spectra for this
type of excitation are plotted in fig.(\ref{fig:so15}), which is 
a two-parameter family of excitation.
The states with negative momentum are just obtained by shifting the 
$\{h_j\}$ in (\ref{eq:QN15}) to the left by one unit.

\begin{figure}
\epsfclipoff
\fboxsep=0pt
\setlength{\unitlength}{1mm}
\begin{picture}(150,180)(0,0)
\linethickness{1pt}
\epsfysize=8cm
\put(18,88){{\epsffile{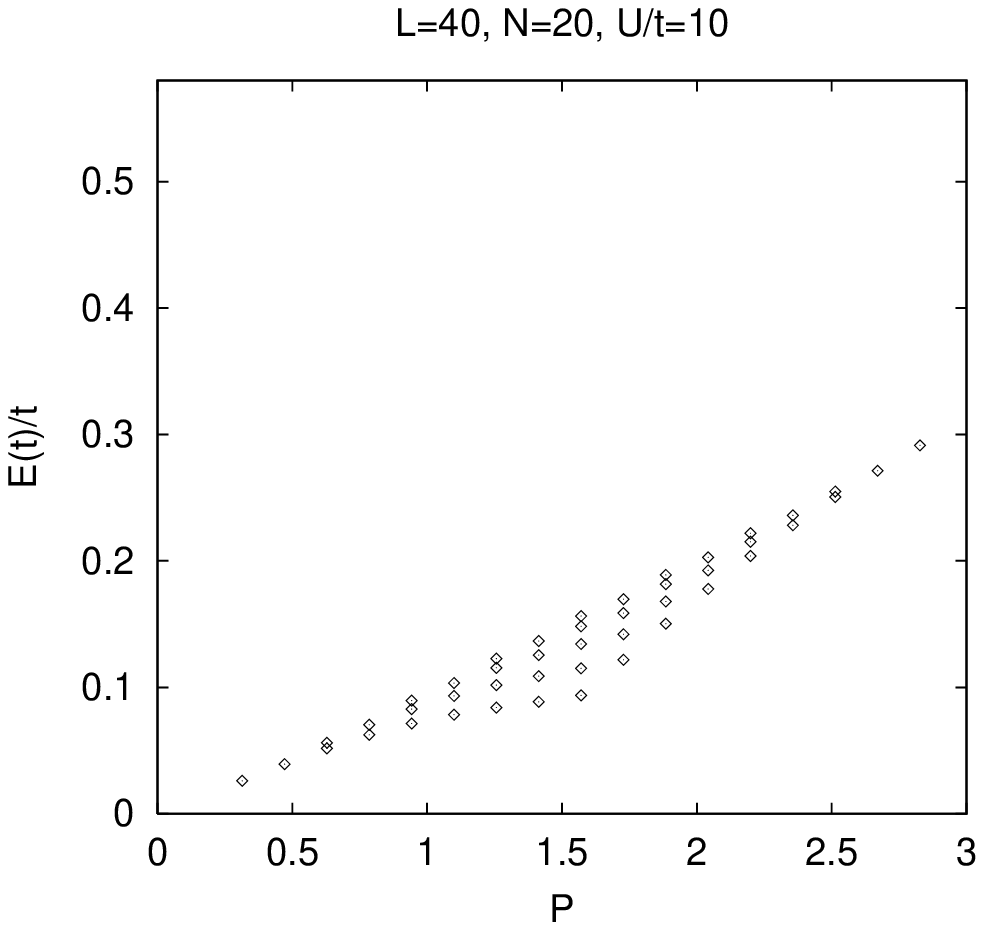}}}
\epsfysize=8cm
\put(18,1){{\epsffile{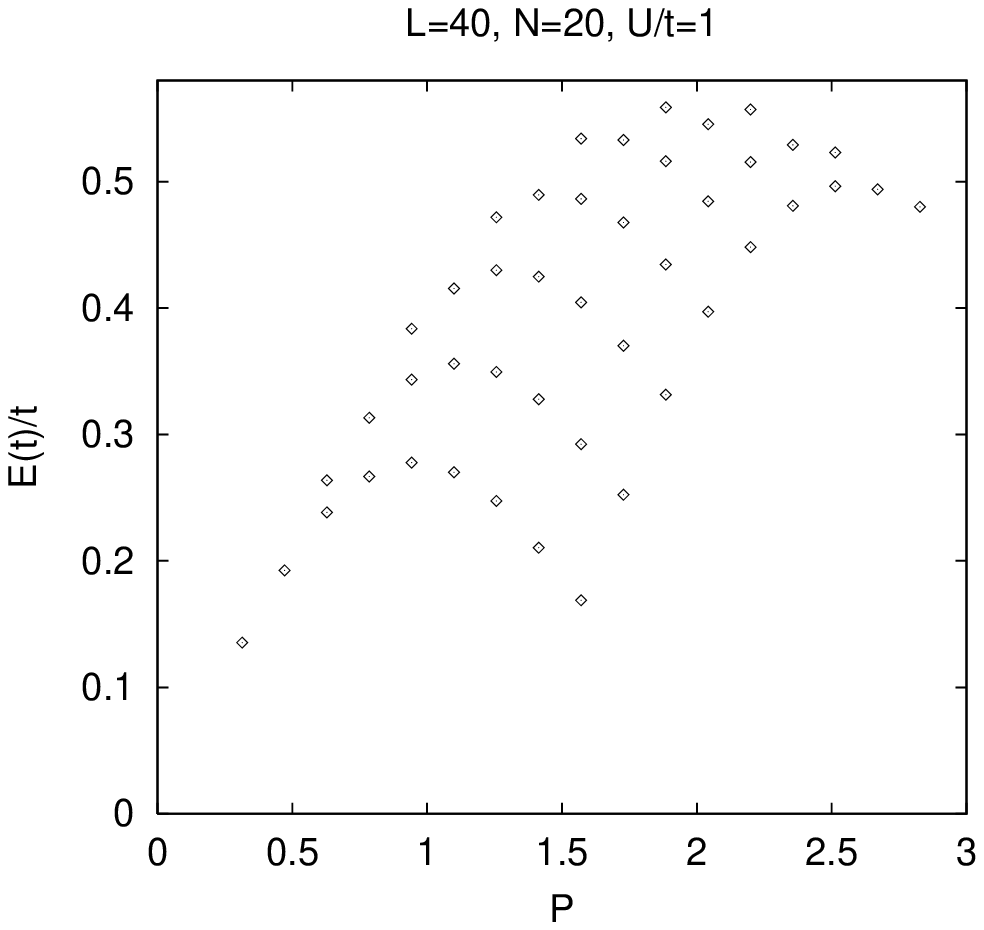}}}
\end{picture}
\vspace{12mm}
\caption{20-fold excitation spectrum for the model
of 40 sites with 20 electrons. Some of points in the upper figure
are almost overlapped.
The overall structure of the spectra changes much with respect to
the correlation strength.
The lowest excitation energy as well as the pattern 
are raised when the correlation strength decreased.}
\label{fig:so20}
\end{figure}

Moving one box from the fourth and one from 
the third row of the Young 
tableau for the ground state to the first and second row, we get
the 20-fold excitation with
$M=3N_0/4-1$, $M'=N_0/2-2$ and $M''=N_0/4-1$. 
The $I$'s and $K$'s that took integer values in the ground state
now take half-integer values; the $h$'s that took half-integer values
now takes integer values and the $J$'s still take half-integer values. 
As a result, two holes in $\{J_\beta\}$ are indispensably appeared,
\begin{eqnarray}
\{h_j\}&=&\{-N_0/2+1, ..., N_0/2-1, N_0/2\},\nonumber\\
\{I_\alpha\}&=&\{-3N_0/8+1,...,3N_0/8-1\},\nonumber\\
\{K_\gamma\}&=&\{-N_0/8+1,...,N_0/8-1\},\nonumber\\
J_1&=&-\frac{N_0-2}{4}+\delta_{1,\beta_1},\nonumber\\
J_\beta &=&J_{\beta-1}+1+\delta_{\beta,\beta_1}+\delta_{\beta,\beta_2},
\label{eq:QN20}
\end{eqnarray}
where $\beta=2,...,N_0/2-2$ and $1\leq\beta_1\leq\beta_2\leq N_0/2$.
Numerical results of the energy-momentum spectra for this
type of excitation, a two-parameter family, 
are given in fig.(\ref{fig:so20}). 
Similarly, the states with negative momentum are just obtained 
by shifting the $\{h_j\}$ in (\ref{eq:QN20}) to the left by one unit.

\begin{figure}
\epsfclipoff
\fboxsep=0pt
\setlength{\unitlength}{1mm}
\begin{picture}(150,180)(0,0)
\linethickness{1pt}
\epsfysize=8cm
\put(18,88){{\epsffile{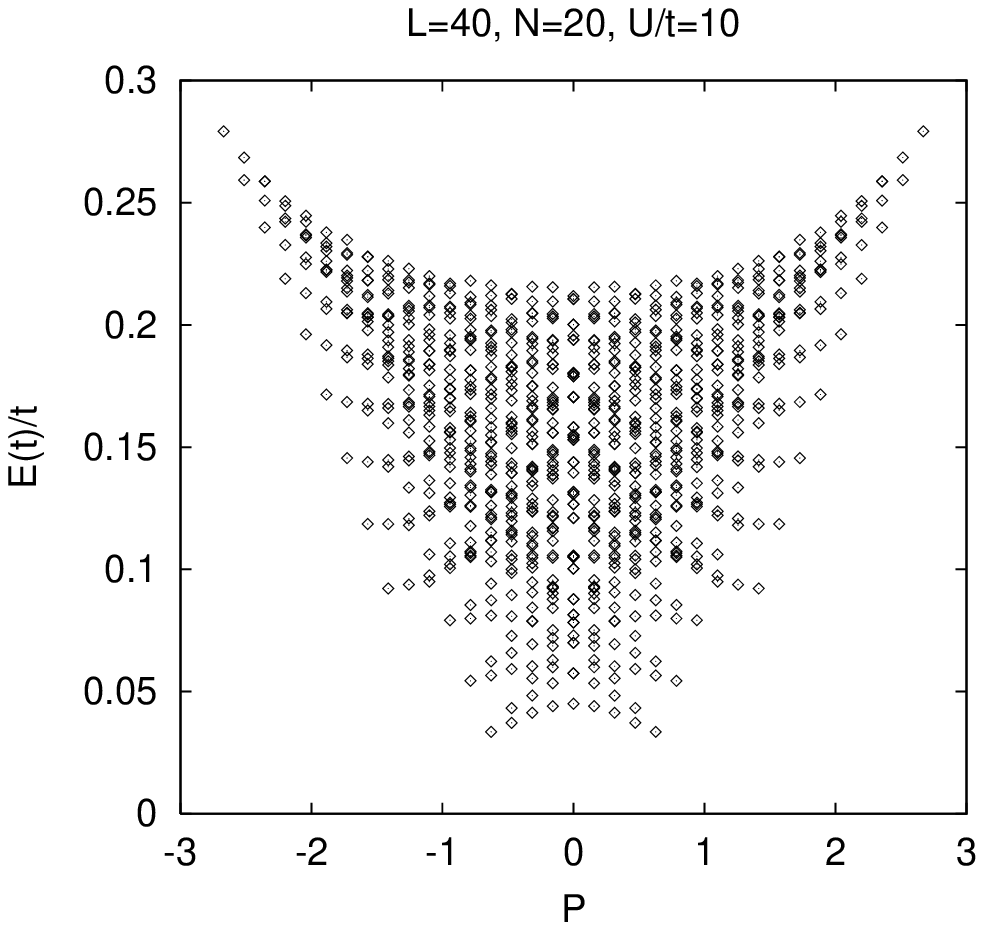}}}
\epsfysize=8cm
\put(18,1){{\epsffile{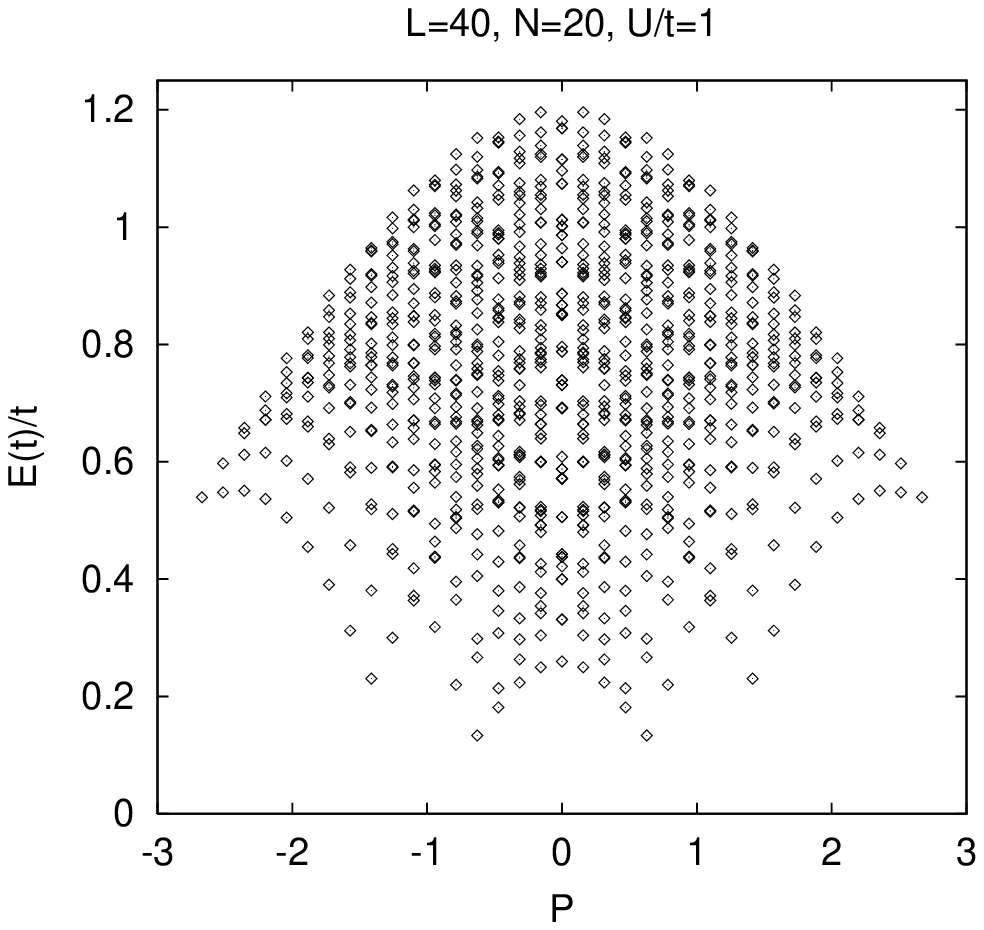}}}
\end{picture}
\vspace{12mm}
\caption{45-fold excitation spectrum for the model
of 40 sites with 20 electrons. }
\label{fig:so45}
\end{figure}

Moving  one box from the fourth and one from the third row of the Young 
tableau for the ground state to the first row, we get the 45-fold
excitation with 
$M=3N_0/4-2$, $M'=N_0/2-2$ and $M''=N_0/4-1$.
This makes the 
$J$'s to take integer values instead of half-integer ones
and $K$'s to take half-integer values instead of integer ones.
In this case, there exist two holes in $\{I_\alpha\}$ 
and one hole in $\{J_\beta\}$, namely
\begin{eqnarray}
\{h_j\}&=&\{-N_0/2+1, ..., N_0/2-1\},\nonumber\\
\{K_\gamma\}&=&\{-N_0/8+1,...,N_0/8-1\},\nonumber\\
I_1&=&-\frac{3N_0-4}{8}+\delta_{1,\alpha_1},\nonumber\\
I_\alpha &=&I_{\alpha-1}+1+\delta_{\alpha,\alpha_1}
                 +\delta_{\alpha,\alpha_2},\nonumber\\
J_1 &=&-\frac{N_0}{4}+1+\delta_{1,\beta_1},\nonumber\\
J_\beta &=& J_{\beta-1}+1+\delta_{\beta,\beta_1}\,\,\,
(\beta=2,...,\frac{N_0}{2}-2),
\label{eq:QN45}
\end{eqnarray}
where $\alpha=2,...,3N_0/4-2$, $1\leq\alpha_1<\alpha_2\leq 3N_0/4$
and $1\leq\beta_1\leq N_0/2-1$. Numerical results are plotted 
in fig.(\ref{fig:so45}). It is a three-parameter family of excitation.

\begin{figure}
\epsfclipoff
\fboxsep=0pt
\setlength{\unitlength}{1mm}
\begin{picture}(150,180)(0,0)
\linethickness{1pt}
\epsfysize=8cm
\put(18,88){{\epsffile{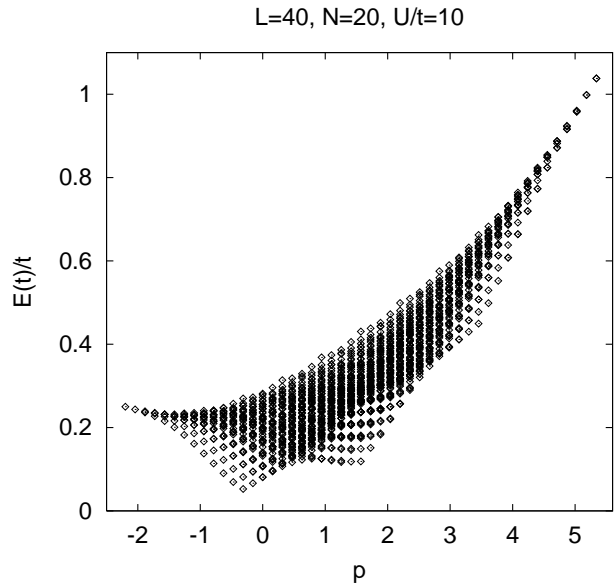}}}
\epsfysize=8cm
\put(18,1){{\epsffile{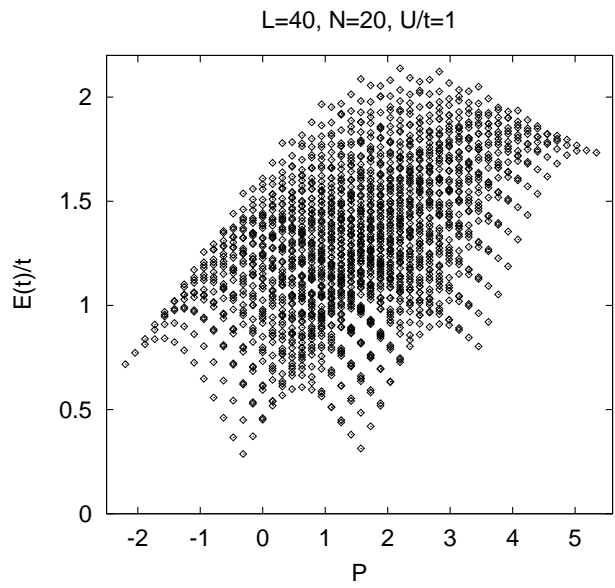}}}
\end{picture}
\vspace{12mm}
\caption{35-fold excitation spectrum for the model
of 40 sites with 20 electrons. The complete spectra include the 
other part which is just the mirror image of the above pattern and
is not plotted out.}
\label{fig:so35}
\end{figure}

Taking out one box respectively from the second, 
the third and the fourth row
and putting them together
on the first row of the Young tableau, we
have a 35-fold excitation. 
Due to $M=3N_0/4-3$, $M'=N_0/2-2$ and $M''=N_0/4-1$,
we have four holes in $\{I_\alpha\}$, accordingly,
\begin{eqnarray}
\{h_j\}&=&\{-N_0/2+1, ..., N_0/2-1, N_0/2\}, \nonumber\\
\{J_\beta\}&=&\{-(N_0-6)/4, ..., (N_0-6)/4\}, \nonumber\\
\{K_\gamma\}&=&\{-N_0/8+1, ..., N_0/8-1\},\nonumber\\
I_1&=&-\frac{3N_0}{4}+\delta_{1,\alpha_1},\nonumber\\
I_\alpha &=&I_{\alpha-1}+1+\sum_{i=1}^4\delta_{\alpha,\alpha_i}\,\,\,
  (\alpha=2,...,3N_0/4-3),
\label{eq:QN35}
\end{eqnarray}
where $1\leq\beta_1<\beta_2<\beta_3<\beta_4\leq 3N_0/4+1$.
The numerical results are plotted in fig.(\ref{fig:so35}),
where we did not plot the pattern obtained by shifting $\{h_j\}$ in
(\ref{eq:QN35}) to the left by one unit, which is just the mirror
image of the plotted pattern. This is a four-parameter family of 
excitation. 

\subsection{Adding particles}

\begin{figure}
\epsfclipoff
\fboxsep=0pt
\setlength{\unitlength}{1mm}
\begin{picture}(150,180)(0,0)
\linethickness{1pt}
\epsfysize=8cm
\put(18,88){{\epsffile{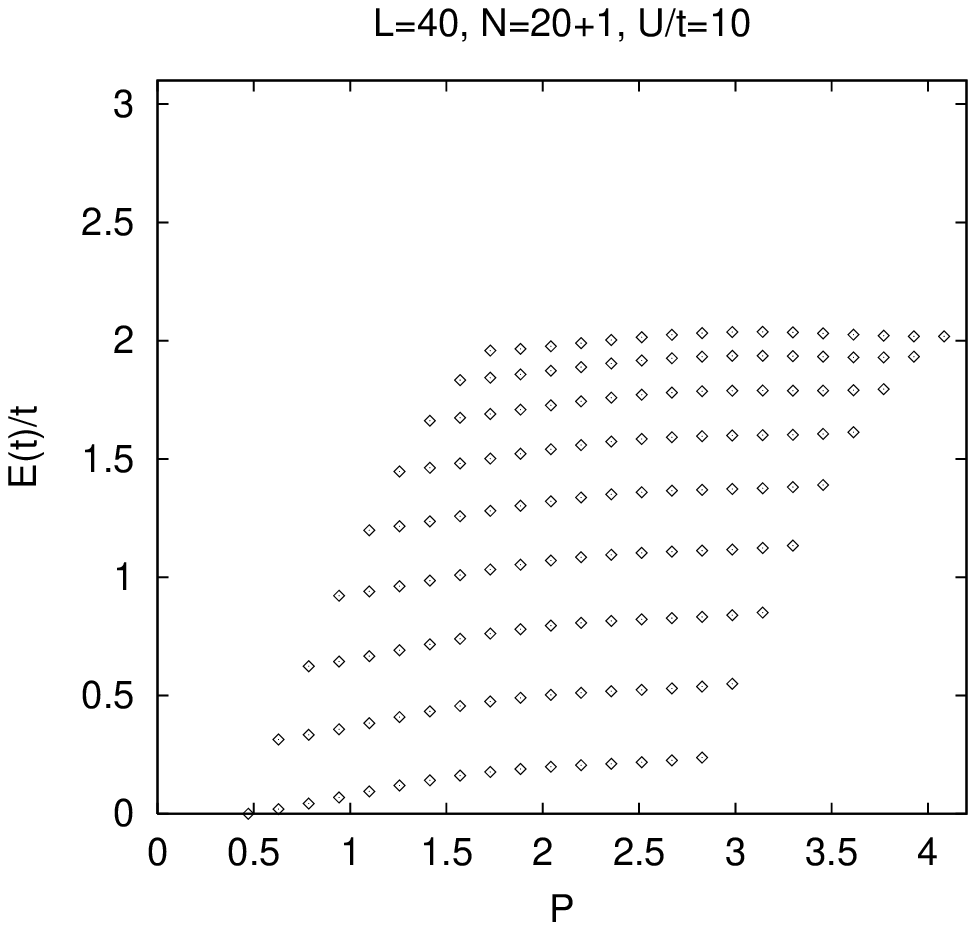}}}
\epsfysize=8cm
\put(18,1){{\epsffile{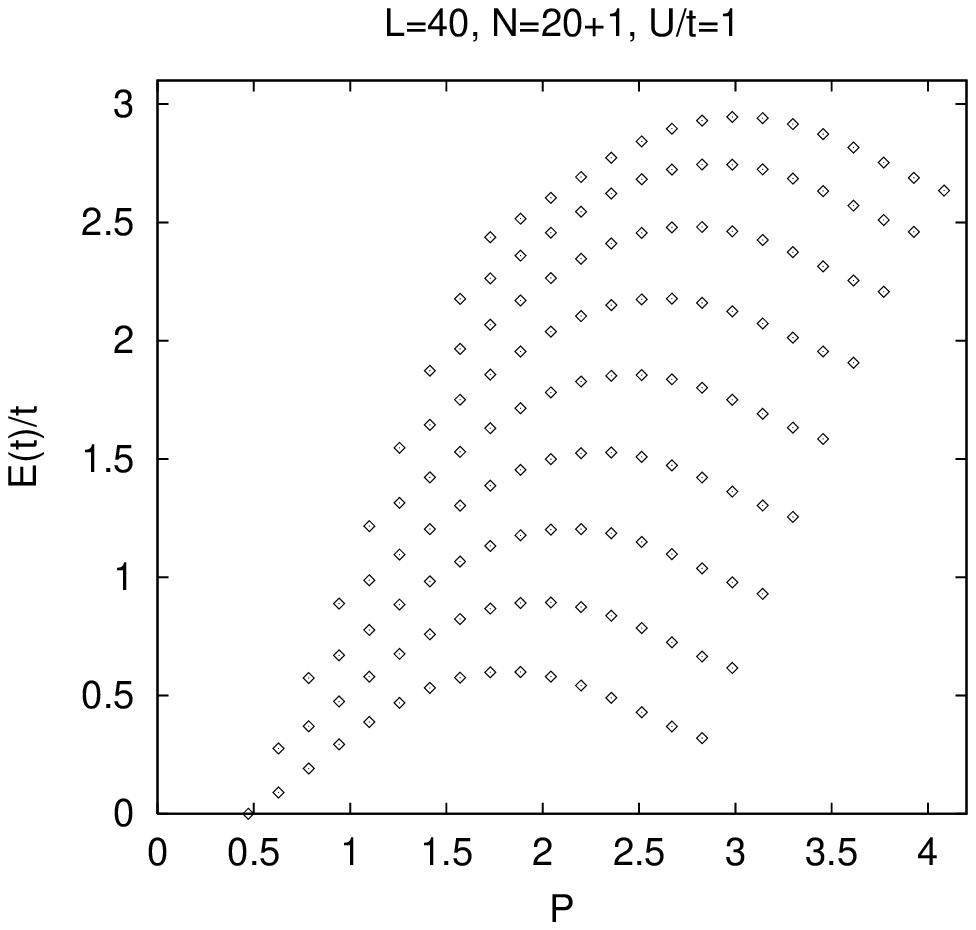}}}
\end{picture}
\vspace{12mm}
\caption{Excitation spectrum for one particle added into the 
model of 40 sites with 20 electrons. The zero energy 
corresponds to the 21 electron ground state.}
\label{fig:add1}
\end{figure}

If the number of electrons are $N_0+1$, $N_0+2$ or $N_0+3$, the corresponding
states can be obtained by adding one, two or three particles into the system
of $N_0$ electrons. 
Adding one particle to the $N_0$ ground state and leaving $M$, $M'$ and $M''$
unchanged, the $h$'s, $I$'s and $J$'s are half-odd integers but $K$'s
are integers. Comparing to that for the non-degenerate ground state 
of $N_0=4n$, there are now $3N_0/4+1$ allowed values for the $3N_0/4$
distinct $I$'s. So there is always a ``hole'' in the $I$'s sequence, namely,
\begin{eqnarray}
\{h_j\}&=&\{-(N_0-1)/2,...,(N_0-1)/2, h_0\}, \nonumber\\
I_1&=&3N_0/8+\delta_{1,\alpha_1}, \nonumber\\
I_\alpha&=&I_{\alpha-1}+1+\delta_{\alpha,\alpha_1} \, 
  (\alpha=2,..., 3N_0/4),
\label{eq:QNadd1}
\end{eqnarray}
where $1\leq\alpha_1\leq3N_0/4+1$. The $J$'s and $K$'s sequences 
are the same as those in (\ref{eq:QNground}). The numerical results for
$h_0>0$ are plotted in fig.(\ref{fig:add1}), where the zero energy 
corresponds to the ground state for $N=N_0+1$. 
The spectra with negative momentum are obtained by using $h_0<0$.
It is easy to know by evaluating eq.(\ref{eq:HWvector}) that
each point in the figure represents a quadruplet.

\begin{figure}
\epsfclipoff
\fboxsep=0pt
\setlength{\unitlength}{1mm}
\begin{picture}(150,180)(0,0)
\linethickness{1pt}
\epsfysize=8cm
\put(18,88){{\epsffile{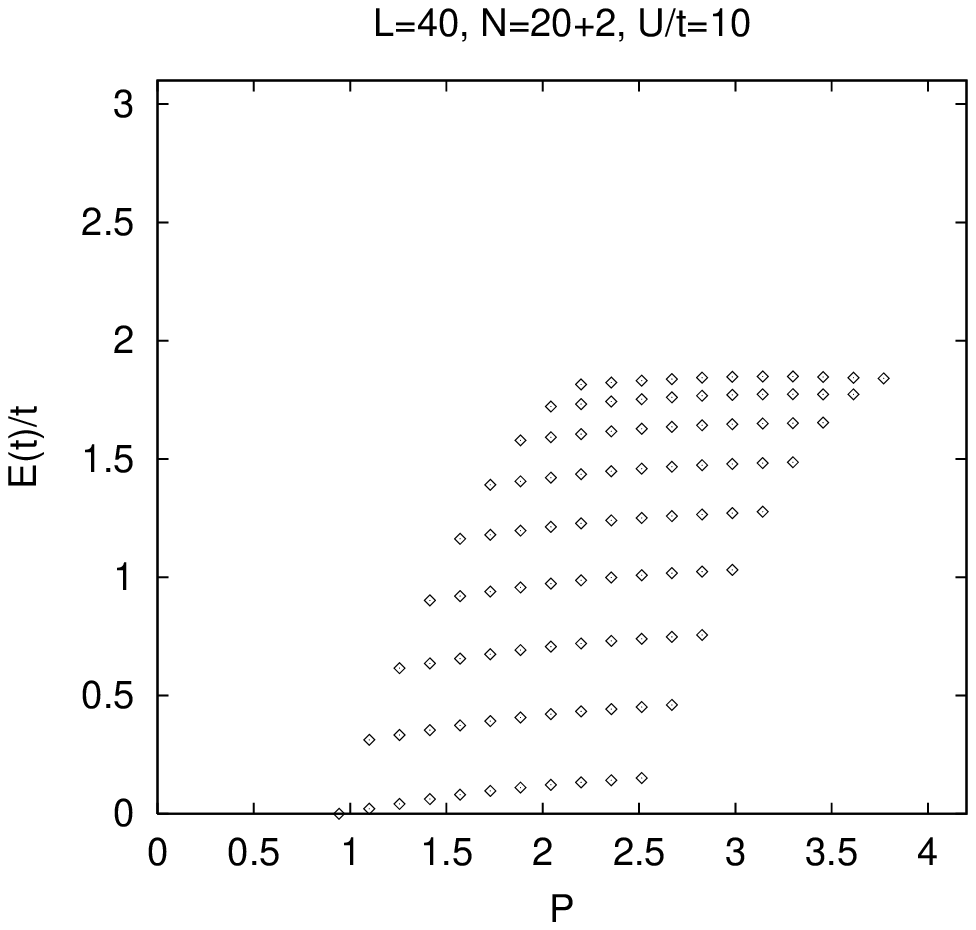}}}
\epsfysize=8cm
\put(18,1){{\epsffile{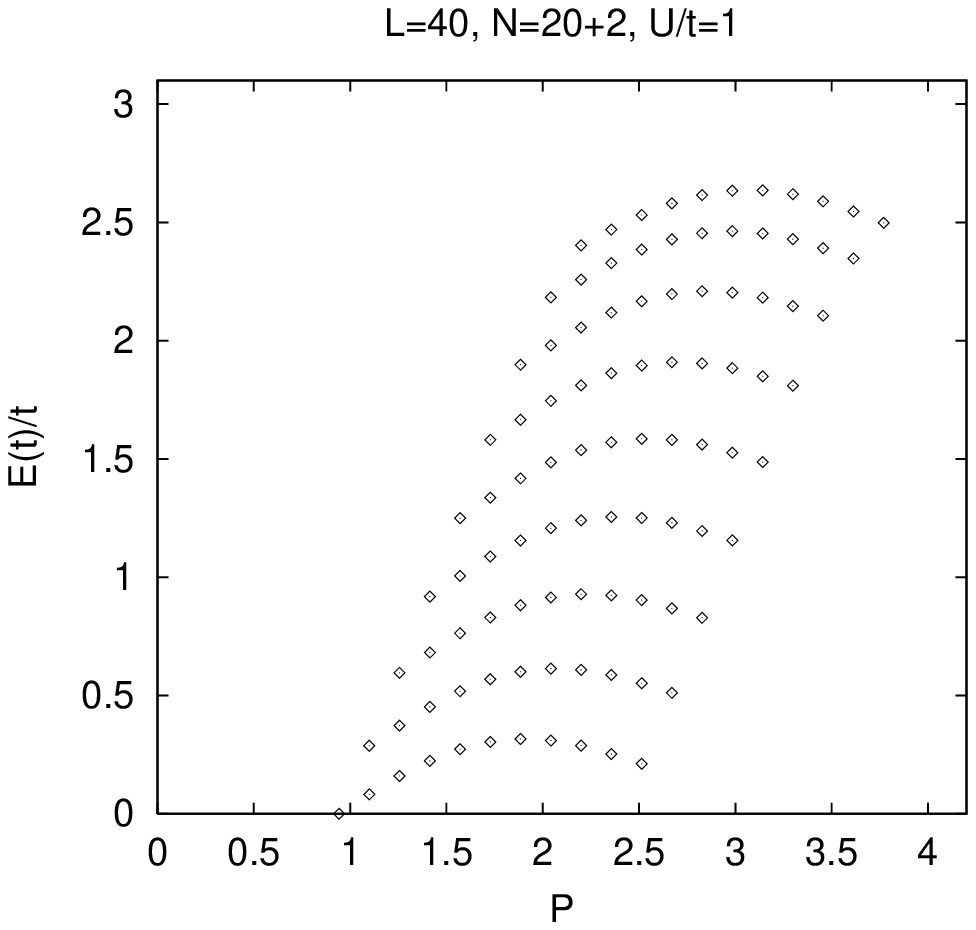}}}
\end{picture}
\vspace{12mm}
\caption{Excitation spectrum for two particles added into the 
model of 40 sites with 20 electrons. The zero energy 
corresponds to the 22 electron ground state.}
\label{fig:add2}
\end{figure}

Adding two particles to the $N_0$ particle ground state, 
we have $N=N_0+2$, $M=3N_0/4+1$, $M'=N_0/2$ and $M''=N_0/4$ 
for the low-energy states. 
This requires $h$'s, $J$'s and $K$'s to take integer values but
$I$'s to take half of odd integer values. Referring to (\ref{eq:QNground})
we know that there must be a ``hole'' in the $J$'s sequence, consequently,
\begin{eqnarray}
\{h_j\}&=&\{-N_0/2,...,N_0/2, h_0\}, \nonumber\\
\{I_\alpha\}&=&\{-3N_0/8,...,3N_0/8\},\nonumber\\
J_1&=&-\frac{N_0}{4}+\delta_{1,\beta_1},\nonumber\\
J_\beta&=&J_{\beta-1}+1+\delta_{\beta,\beta_1}\,
(\beta=2,...,N_0/2),
\label{eq:QNadd2}
\end{eqnarray}
where $1\leq\beta_1\leq N_0/2+1$. The $K$'s sequence is the same as that
in (\ref{eq:QNground}). The numerical results are plotted in 
fig.(\ref{fig:add2}).

\begin{figure}
\epsfclipoff
\fboxsep=0pt
\setlength{\unitlength}{1mm}
\begin{picture}(150,180)(0,0)
\linethickness{1pt}
\epsfysize=8cm
\put(18,88){{\epsffile{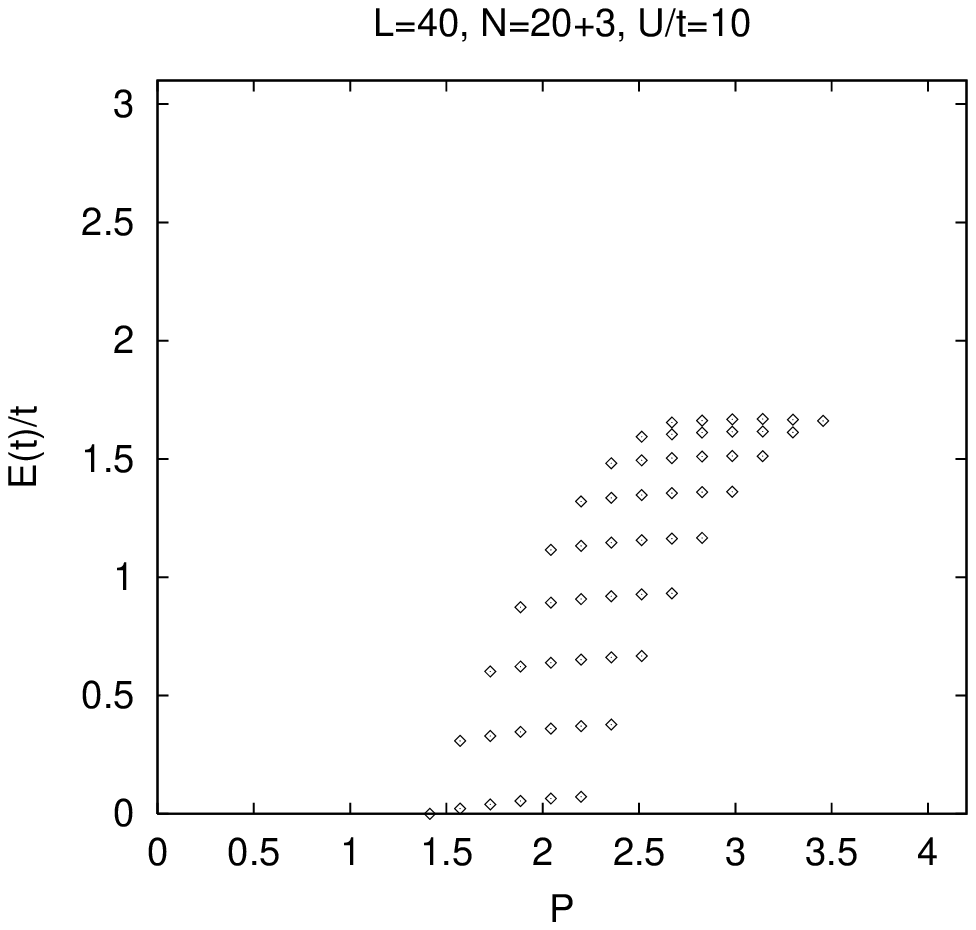}}}
\epsfysize=8cm
\put(18,1){{\epsffile{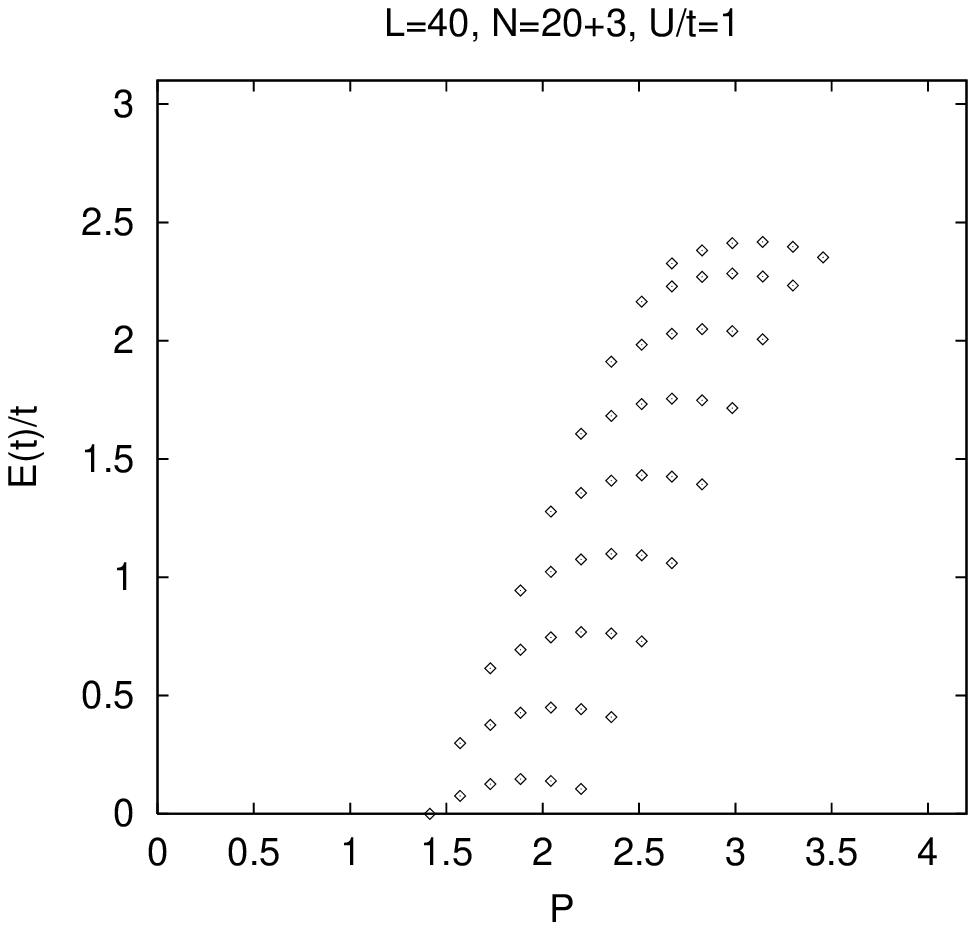}}}
\end{picture}
\vspace{12mm}
\caption{Excitation spectrum for three particles added into the 
model of 40 sites with 20 electrons. The zero energy 
corresponds to the 23 electron ground state.}
\label{fig:add3}
\end{figure}

Adding three particles to the $N_0$ ground state, we have
$N=N_0+3$, $M=3N_0/4+2$, $M'=N_0/2+1$ and $M''=N_0/4$ for low-energy states.
This requires $h$'s and $K$'s to be half-odd integers but
$I$'s and $J$'s  to be integers, and then 
\begin{eqnarray}
\{h_j\}&=&\{-(N_0+1)/2,...,(N_0+1)/2, h_0\}, \nonumber\\
\{I_\alpha\}&=&\{-(3N_0+4)/8,...,(3N_0+4)/8\},\nonumber\\
\{J_\beta\}&=&\{-N_0/8,...,N_0/4\},\nonumber\\
K_1&=&-\frac{N_0}{8}+\delta_{1,\gamma_1},\nonumber\\
K_\gamma&=&K_{\gamma-1}+1+\delta_{\gamma,\gamma_1}\,
(\gamma=2,...,N_0/4),
\label{eq:QNadd3}
\end{eqnarray}
where $1\leq\gamma_1\leq N_0/4+1$. 
The numerical results are plotted in fig.(\ref{fig:add3}).

\section{Special cases}\label{sec:special}

For some special limiting cases,
one is able to obtain several interesting conclusions
from the Bethe-ansatz equation without solving it directly\cite{Schulz}.
In the following, we consider the weak coupling and strong coupling
respectively.

\subsection{Weak coupling}

Because $\Theta_n(x)\rightarrow\pi{\rm sign}(x)$ for $c\rightarrow 0$,
the Bethe-ansatz equation (\ref{eq:BAE}) becomes
\twobeone
\begin{eqnarray}
k_j+\frac{\pi}{L}\sum_{\alpha=1}^M{\rm sgn}(\sin k_j-\lambda_\alpha)
   &=&\frac{2\pi}{L}h_j, 
       \nonumber\\
\sum_{l=1}^{N}{\rm sgn}(\lambda_\alpha-\sin k_l)
  -\sum_{\alpha'=1}^{M}{\rm sgn}(\lambda_\alpha-\lambda_{\alpha'})
   +\sum_{\beta=1}^{M'}{\rm sgn}(\lambda_\alpha-\mu_\beta)
     &=&2 I_\alpha,     
        \nonumber\\
\sum_{\alpha=1}^{M}{\rm sgn}(\mu_\beta-\lambda_\alpha)
  -\sum_{\beta'=1}^{M'}{\rm sgn}(\mu_\beta-\mu_{\beta'})
    +\sum_{\gamma=1}^{M''}{\rm sgn}(\mu_\beta-\nu_\gamma)
      &=&2 J_\beta,
        \nonumber\\
\sum_{\beta=1}^{M'}{\rm sgn}(\nu_\gamma-\mu_\beta)
  -\sum_{\gamma'=1}^{M''}{\rm sgn}(\nu_\gamma-\nu_{\gamma'})
    &=&2 K_\gamma.
\label{eq:BAEweak}
\end{eqnarray}
\onebetwo
Without loss of generality, the $\gamma$ label can be so chosen that
$K_\gamma$ is arranged in an increasing order. Then the fourth
equation of (\ref{eq:BAEweak}) becomes
\begin{eqnarray}
\sum_{\mu=1}^{M'}{\rm sgn}(\nu_\gamma-\mu_\beta)
 =2K_\gamma+2\gamma-M''-1.
\label{eq:BAE4weak}
\end{eqnarray}
Because $|K_\gamma|\leq(M''-1)/2$ and $M''\leq M'/2$ (restricted by the 
Young tableau), the minimum value of right-hand side of (\ref{eq:BAE4weak})
is $-M'+2$. This means that the smallest $\mu_\beta$ is smaller than
the smallest $\nu_\gamma$. 
From eq.(\ref{eq:BAE4weak}) we easily derive
\begin{eqnarray}
\sum_{\beta=1}^{M'}[{\rm sgn}(\nu_{\gamma+1}-\mu_\beta)
-{\rm sgn}(\nu_{\gamma}-\mu_\beta)] \nonumber\\
=2(K_{\gamma+1}-K_\gamma +1).
\end{eqnarray}
Obviously, if $K_{\gamma+1}-K_\gamma=m''$, there must exist exactly
$m''+1$ solutions $\mu_\beta$ satisfying 
$\nu_\gamma<\mu_\beta<\nu_{\gamma+1}$.

Again from the third equation of (\ref{eq:BAEweak}), we have
\begin{eqnarray}
\sum_{\alpha=1}^{M}[{\rm sgn}(\mu_{\beta+1}-\lambda_\alpha)
  -{\rm sgn}(\mu_\beta-\lambda_\alpha)]
    =2J_{\beta+1}-2J_\beta \nonumber\\
 + 2-\sum_{\gamma=1}^{M''}[{\rm sgn}(\mu_{\beta+1}-\nu_\gamma)
  -{\rm sgn}(\mu_\beta-\nu_\gamma)]. 
\label{eq:BAE3weak}
\end{eqnarray}
Obviously, if there is a $\nu_\gamma$ such that 
$\mu_\beta<\nu_\gamma<\mu_{\beta+1}$ then the right hand side of 
(\ref{eq:BAE3weak}) equals $2(J_{\beta+1}-J_\beta)$,
otherwise it equals  $2(J_{\beta+1}-J_\beta +1)$.
Proceeding the same strategy to the second equation
and the first equation in (\ref{eq:BAEweak}) successively, we obtain
a sequence of relations that are summarized as follows.

(i) If $K_{\gamma+1}-K_\gamma=m''$, there exist exactly $m''+1$ solutions
$\mu_\beta$ satisfying $\nu_\gamma<\mu_\beta<\nu_{\gamma+1}$.

(ii) For $J_{\beta+1}-J_\beta=m'$, if there exists a $\nu_\gamma$ 
satisfying $\mu_\beta<\nu_\gamma<\mu_{\beta+1}$, there will be $m'$ 
$\lambda$'s such that $\mu_\beta<\lambda_\alpha<\mu_{\beta+1}$;
otherwise if there is no $\nu_\gamma$ satisfying that, then there will
be $m'+1$ $\lambda$'s such that $\mu_\beta<\lambda_\alpha<\mu_{\beta+1}$.

(iii) For $I_{\alpha+1}-J_\alpha=m$, if there is a $\mu_\beta$ 
satisfying $\lambda_\alpha<\mu_\beta<\lambda_{\alpha+1}$, there will exist 
$m$ $\sin k_l$'s such that $\lambda_\alpha<\sin k_l<\lambda_{\alpha+1}$;
otherwise there will be $m+1$ $\sin k_l$'s such that 
$\lambda_\alpha<\sin k_l<\lambda_{\alpha+1}$

(iv) For $h_{j+1}-h_j=n'$, if there exists such a $\lambda_\alpha$
that $\sin k_j<\lambda_\alpha<\sin k_{j+1}$, then we will have 
$k_{j+1}-k_j=2\pi(n'-1)/L$; otherwise we will have
$k_{j+1}-k_j=2\pi n'/L$.

Apply these items to the ground state $n'=m=m'=m''=1$,
we conclude that the sequence $\{k_j\}$ is divided into groups 
with four successive values in each group. Between each pair
$(k_{4j-3}, k_{4j-2})$, $(k_{4j-2}, k_{4j-1})$ or $(k_{4j-1}, k_{4j})$ 
in the same group there is one 
$\lambda$, so totally there are three $\lambda$'s
in each group. Furthermore there are two $\mu$'s each lying
in between the adjacent $\lambda$'s. And between these two
$\mu$'s there is always a $\nu$.

\subsection{Strong coupling}

For the strong coupling limit case, $c\rightarrow\infty$ 
(i.e. large $U$ limit), 
the ration $\sin k_j/c$ in the Bethe-ansatz equation is 
neglectable. The first equation of (\ref{eq:BAE}) then becomes
\begin{eqnarray}
k_j&=&\frac{2\pi}{L}n_j-\frac{1}{L}P_H,\nonumber\\
P_H&=&\sum_{\alpha=1}^M\left[\pi-2\tan^{-1}(2\lambda_\alpha/c)\right],
\label{eq:HeisenbergP}
\end{eqnarray}
where $n_j$ are integers.
Obviously, the $P_H$ is just the momentum of spin-orbital excitations
in the SU(4) Heisenberg chain\cite{LiMSZ99,Sutherland75}. The other
three Bethe-ansatz equations turned to be those known for the
SU(4) Heisenberg chain for the scaled rapidities, 
$\lambda_\alpha/c$, $\mu_\beta/c$ and $\nu_\gamma/c$.
Equation (\ref{eq:HeisenbergP}) indicates that the allowed quasi-momenta
$k$'s in the strong coupling limit are quantized in units of $2\pi/L$,
just like ``spinless'' noninteracting fermions. This is because the
double occupancy is forbidden by the strong repulsive on-site
coupling. The allowed quasimomenta due to periodic boundary condition are 
determined by spin-orbital momentum $P_H$.
Particularly, for $N=4n$ with $n=odd$, the ground-state spin-orbital
momentum $P_H$ is an odd multiple of $\pi$ \cite{LiMSZ99} 
so that the allowed $k$'s are half odd integers multiplied by 
$2\pi/L$. Therefore, the ground state for $N=4n$ ($n=odd$) is uniquely
determined, i.e., it is nondegenerate. 
This is different from the noninteracting case in which the allowed
$k$'s are always integers multiplied by $2\pi/L$ for any $N$ and for each
$k$ there are four states because of the spin-orbital degree of freedom.
 
\section{Thermodynamics limit}\label{sec:thermodynamics}

Replacing $k_j$, $\lambda_\alpha$, $\mu_\beta$ and $\nu_\gamma$ 
in eq. (\ref{eq:logBAE})
by continuous variables $k$, $\lambda$, $\mu$, and $\nu$ 
but keeping the summation still over the solution set of these roots,
we can consider the quantum
numbers $h_j$, $I_\alpha$, $J_\beta$, and $K_\gamma$ as functions 
$h(k)$, $I(\lambda)$, $J(\mu)$, and $K(\nu)$ given by 
eq. (\ref{eq:logBAE}). Take $I(\lambda)$ as an example, when
$I(\lambda)$  passes through one of the quantum numbers
$I_j$, the corresponding $\lambda$ is equal to
one of the roots $\lambda_j$, similarly for $J(\mu)$, $K(\nu)$, or $h(k)$. 
However, there may exist some integers or half-integers for which the
corresponding $\lambda$ ($\mu$, $\nu$, or $k$) is not in the set of roots.
Such a situation is conventionally referred as a ``hole''. 
In the thermodynamics limit, 
$N\rightarrow\infty$, $L\rightarrow\infty$, but $N/L$ kept finite, 
we may introduce the density of real roots and
the density of holes (indicated by a subscript $h$), 

\begin{eqnarray}
\rho(k)+\rho_h(k)=(1/L)dh(k)/dk,
     \nonumber\\
\sigma(\lambda)+\sigma_h(\lambda) = (1/L)dI(\lambda)/d\lambda,
     \nonumber\\
\omega(\mu)+\omega_h(\mu) =(1/L)dJ(\mu)/d\mu, 
     \nonumber\\ 
\tau(\nu)+\tau_h(\nu) =(1/L)dK(\nu)/d\nu.
     \nonumber
\end{eqnarray}
By replacing the summations by integrals, for example,
\begin{eqnarray}
\lim_{L\rightarrow\infty}\frac{1}{L}\sum_{l=1}^N f(k_l)
=\int_{-Q}^{Q}dk\rho(k)f(k),
            \nonumber\\
\lim_{L\rightarrow\infty}\frac{1}{L}\sum_{\alpha=1}^M g(\lambda_\alpha)
=\int^B_{-B}d\lambda\sigma(\lambda)g(\lambda),
\end{eqnarray}
and so forth, eq.(\ref{eq:logBAE}) gives rise to the following
coupled integral equations,

\twobeone
\begin{eqnarray}
\rho(k)+\rho^{(o)}(k)
&=&\frac{1}{2\pi}
 -\cos k\int_{-B}^{B}K_{-1/2}(\sin k-\lambda)\sigma(\lambda)d\lambda,
       \nonumber\\
\sigma(\lambda)+\sigma^{(o)}(\lambda)
 &=&-\int_{-Q}^{Q}K_{-1/2}(\lambda-\sin k)\rho(k)dk
   -\int_{-B}^{B}K_1(\lambda-\lambda')\sigma(\lambda')d\lambda'
    -\int_{-B'}^{B'}K_{-1/2}(\lambda-\mu)\omega(\mu)d\mu,
      \nonumber\\
\omega(\mu)+\omega^{(o)}(\mu) 
&=&-\int_{-B}^{B}K_{-1/2}(\mu-\lambda)\sigma(\lambda)d\lambda
   -\int_{-B'}^{B'}K_1(\mu-\mu')\omega(\mu')d\mu'
    -\int_{-B''}^{B''}K_{-1/2}(\mu-\nu)\tau(\nu)d\nu,
      \nonumber\\ 
\tau(\nu)+\tau^{(o)}(\nu)  
&=&-\int_{-B'}^{B'}K_{-1/2}(\nu-\mu)\omega(\mu)d\mu
    -\int_{-B''}^{B''}K_1(\nu-\nu')\tau(\nu')d\nu',
\label{eq:densityEqs}
\end{eqnarray}
\onebetwo
where $K_n(x):=\pi^{-1}nc/(n^2c^2+x^2)$,
and $Q$, $B$, $B'$, and $B''$
in the definite integrals should be 
determined self-consistently by 
\begin{eqnarray}
\frac{N}{L}&=&\int^{Q}_{-Q}\rho(k)dk,\nonumber\\  
\frac{M}{L}&=&\int^B_{-B}\sigma(\lambda)d\lambda,\nonumber\\ 
\frac{M'}{L}&=&\int^{B'}_{-B'}\omega(\mu)d\mu,\nonumber\\ 
\frac{M''}{L}&=&\int^{B''}_{-B''}\tau(\nu)d\nu.
\end{eqnarray}
They hold for the case in the absence of the complex roots. In the 
presence of complex roots, however, it has variants.
In eq.(\ref{eq:densityEqs}) 
we denoted the inhomogeneous terms by $\rho^{(o)}$, $\sigma^{(o)}$, 
$\omega^{(o)}$ and $\tau^{(o)}$,
which not only stand for the densities of holes $\rho_h$, $\sigma_h$ 
etc., but also the contributions from complex roots, two-strings. 

Once the density $\rho(k)$ is solved from eq.(\ref{eq:densityEqs}), 
we have the $z$-components of the total spin and the total orbital
\begin{eqnarray}
\frac{S^z_{tot}}{L}
  &=&\frac{1}{2}\int_{-Q}^{Q}\rho(k)dk
   +\int_{-B'}^{B'}\omega(\mu)d\mu
     \nonumber\\
\,&\,&-\int_{-B}^{B}\sigma(\lambda)d\lambda 
 -\int_{-B''}^{B''}\tau(\nu)d\nu,
   \nonumber\\
\frac{T^z_{tot}}{L}
  &=&\frac{1}{2}\int_{-Q}^{Q}\rho(k)dk
   -\int_{-B'}^{B'}\omega(\mu)d\mu.
\label{eq:totalspin}
\end{eqnarray}
This is useful for a correct calculation of magnetizations.
The energy is given by
\begin{equation}
\frac{E}{L}=-2t\int_{-Q}^{Q}\cos k \rho(k)dk.
\label{eq:Eint}
\end{equation}
The highest weight vector that characterizes the corresponding 
representation is given by
\begin{eqnarray}
w_1 &=& L\int^{Q}_{-Q}\rho(k)dk
  -2L\int^B_{-B}\sigma(\lambda)d\lambda
   +L\int^{B'}_{-B'}\omega(\mu)d\mu,
    \nonumber\\
w_2 &=& L\int^B_{-B}\sigma(\lambda)d\lambda
   -2L\int^{B'}_{-B'}\omega(\mu)d\mu  
     +L\int_{-B''}^{B''}\tau(\nu)d\nu,
      \nonumber\\
w_3 &=& L\int^{B'}_{-B'}\omega(\mu)d\mu
  -2L\int_{-B''}^{B''}\tau(\nu)d\nu.
\label{eq:HWintegral}
\end{eqnarray}

\subsection{Ground state properties}

The ground state of the present model is a Fermi
sea described by $\rho_0(k)$, which is the distribution 
function of charge with respect to momentum $k$. 
The $\tau_0(\nu)$ describes the distribution 
of states with spin down and orbital bottom in the $\nu$-rapidity
space. The $\omega_0(\mu)$ represents the distribution of 
either the state with spin up while orbital bottom or
that with spin down while orbital bottom in the $\mu$-rapidity space.  
The $\sigma_0(\lambda)$, however, stands  for  the distribution 
of either state $\braket{2}$, $\braket{3}$ or $\braket{4}$ in
the $\lambda$-rapidity space.
These distribution functions satisfy 
(\ref{eq:densityEqs}) with $B=B'=B''=\infty$ and no holes, i.e., 
$\rho^{(o)}=0$, $\sigma^{(o)}=0$, $\omega^{(o)}=0$,  
$\tau^{(o)}=0$.
By making Fourier transform to the second till  the fourth equation
of eq.(\ref{eq:densityEqs}), we have
\begin{eqnarray}
\tilde{\sigma}(q)&=&\frac{1}{\sqrt{2\pi}}\int_{k_F}^{k_F}
  e^{-c|q|/2+iq\sin k}\rho(k)dk
    \nonumber\\[2mm]
 \,&\,&-\tilde{\sigma}(q)e^{-c|q|}+\tilde{\omega}(q)e^{-c|q|/2},
     \nonumber\\[3mm]
\tilde{\omega}(q)&=&\tilde{\sigma}(q)e^{-c|q|/2}
 -\tilde{\omega}(q)e^{-c|q|}+\tilde{\tau}(q)e^{-c|q|/2},
    \nonumber\\[3mm]
\tilde{\tau}(q)&=&\tilde{\omega}(q)e^{-c|q|/2}-\tilde{\tau}(q)e^{-c|q|}.
\label{eq:FourierGS}
\end{eqnarray}
It is not difficult to obtain a single integral equation 
that determines the $\rho_0(k)$
\begin{equation}
\rho_0(k)=\frac{1}{2\pi}
  +\frac{\cos k}{c}\int_{-k_F}^{k_F}
   R_{3/2}\left(\frac{\sin k-sin k'}{c}\right)\rho_0(k')dk',
\label{eq:groundrho}
\end{equation}
where $k_F$ is the Fermi momentum and
\[
R_{n}(x)=\int_{-\infty}^{\infty}\frac{dq}{2\pi}
  \frac{\sinh(nq)}{\sinh(2q)}e^{iqx-|q|/2}.
\]
Once $\rho_0(k)$ is solved, the energy will be evaluated by the integral
$$
E_0/L=-2t\int_{-k_F}^{k_F}\cos k\rho_0(k)dk.
$$ 
Though an explicit expression cannot be obtained 
from eq.(\ref{eq:groundrho}) in the general case, it becomes
easier for a numerical calculation.

It is immediate from eq.(\ref{eq:FourierGS}) that the highest weight
vector (\ref{eq:HWintegral}) of the ground state is a null vector. 
Therefore the ground state is a SU(4) singlet, accordingly, both spin
and orbital are ``anti-ferromagnetic''.

\subsection{ground-state energy for strong coupling}

The ground-state energy can be calculated explicitly at strong
on-site coupling, $c\gg 1$  (i.e., $U\gg t$). 
Because of $(\sin k-\sin k')/c \ll 1$ in this case
and $4\pi R_{3/2}(0)=3\ln2+\pi/2$, eq. (\ref{eq:groundrho}) 
is written out up to the order $O(1/c)$,
\[
\rho_0^{stro}=\frac{1}{2\pi}
  +(3\ln2+\frac{\pi}{2})\frac{\cos k}{4\pi c}\frac{N}{L}.
\]

The Fermi momentum determined from $N/L=\int_{-k_F}^{k_F}\rho_0(k)dk$ is 
\[
k_F=\frac{N}{L}
   \left[\pi-(3\ln2+\frac{\pi}{2})\frac{\sin(\pi N/L)}{2c}\right],
\]
and then the energy is calculated 
\begin{eqnarray}
\frac{E_0}{L}&=&-\frac{t\sin(\pi N/L)}{\pi/2} \nonumber\\
  \,&-&\frac{t^2}{U}(\frac{N}{L})^2(3\ln2+\frac{\pi}{2})
    \left[1-\frac{\sin(2\pi N/L)}{2\pi N/L}\right],
\end{eqnarray}
where $N/L$ is the filling factor. It becomes 
$E_0^{(1/2)}=-Nt^2(6\ln2+\pi)/U$ at half-filling $N=2L$.
At quarter filling $L=N$ we have
\[
\frac{E^{(1/4)}_0}{N}=-2\frac{t^2}{U}(\frac{3}{2}\ln2+\frac{\pi}{4}),
\]
which agrees with the 
result of the SU(4) Heisenberg model \cite{LiMSZ99,Sutherland75} 
for $J=2t^2/U$. 
Because the model is solved under the assumption of excluding 
site occupations of more than two, the results here are not valid
for above half-filling $N>2L$ in which there must exist sites occupied
by three electrons and the Bethe-ansatz wave-function failures
at that point in  configuration space. However, the energy is 
expected to be evaluate by a particle-hole transformation\cite{LiE99},
\begin{equation}
E(N/L,U)/L=E(4-N/L,U)/L+3U(N/L-2).
\label{eq:p-hTransform}
\end{equation}
In the next section and thereafter we will study low-lying 
excitations on the basis of the thermodynamics limit.

\section{Spin-orbital excitations}\label{sec:spin-orbital}

It is convenient to study the excitations by introducing
$\rho(k)=\rho_0(k) +\rho_1(k)/L$,
$\sigma(\lambda)=\sigma_0 (\lambda)+\sigma_1(\lambda)/L$,
$\omega(\mu)=\omega_0(\mu)+\omega_1(\mu)/L$, and 
$\tau(\nu)=\tau_0(\nu)+\tau_1(\nu)/L $
where $\rho_0(k)$, $\sigma_0 (\lambda)$, $\omega_0(\mu)$,
and $\tau_0(\nu)$ satisfy
the same set of integral equations as the ground state did.
The excitation energy up to the order $O(1/L)$ is
\begin{equation}
 \Delta E = -\int_{-Q}^Q dk (2t\cos k + \Lambda)\rho_1 (k),
\label{eq:Excite}
\end{equation}
where $\Lambda$ stands for the chemical potential \cite{Cornelius}.
$Q$ can be replaced by $k_F$ for a large system.
Equation (\ref{eq:Excite}) is valid for both the spin-orbital excitation 
and the charge excitation. The excitation energy
is related to $\rho_1(k)$, which, moreover, should be solved from the 
following coupled integral equations:
\twobeone
\begin{eqnarray}
\rho_1(k)+\rho_1^{(o)}(k)
&=&-\cos k\int_{-\infty}^{\infty}K_{-1/2}(\sin k-\lambda)
      \sigma_1(\lambda)d\lambda,
       \nonumber\\
\sigma_1(\lambda)+\sigma_1^{(o)}(\lambda)
 &=&-\int_{-k_F}^{k_F}K_{-1/2}(\lambda-\sin k)\rho_1(k)dk
   -\int_{-\infty}^{\infty}K_1(\lambda-\lambda')\sigma_1(\lambda')d\lambda'
    -\int_{-\infty}^{\infty}K_{-1/2}(\lambda-\mu)\omega_1(\mu)d\mu,
      \nonumber\\
\omega_1(\mu)+\omega_1^{(o)}(\mu) 
&=&-\int_{-\infty}^{\infty}K_{-1/2}(\mu-\lambda)\sigma_1(\lambda)d\lambda
   -\int_{-\infty}^{\infty}K_1(\mu-\mu')\omega_1(\mu')d\mu'
    -\int_{-\infty}^{\infty}K_{-1/2}(\mu-\nu)\tau_1(\nu)d\nu,
           \nonumber\\ 
\tau_1(\nu)+\tau_1^{(o)}(\nu)  
&=&\int_{-\infty}^{\infty}K_{-1/2}(\nu-\mu)\omega_1(\mu)d\mu
   +\int_{-\infty}^{\infty}K_1(\nu-\nu')\tau_1(\nu')d\nu'.
\label{eq:densityExcite}
\end{eqnarray}
\onebetwo
The limits for the definite integrals are 
the same as that for ground state,
which are valid for the low-lying excitations. 
Beyond low-lying excitations, however, the integration limits $Q$,
$B$, $B'$, and $B''$ should be determined consistently. Here we only consider  
low-lying excitations.

In order to consider the excitations above the singlet ground state,
we must  analyze the decomposition of the direct
product of the SU(4) fundamental representation for $N=4n$. 
Using the Young tableau, we can obtain that  
the decomposition gives rise to  a direct sum of a
series of irreducible representations, i.e., 
$(0,0,0)$, $(1,0,1)$, $(0,2,0)$,
$(2,1,0)$, $(4,0,0)$ $\it{etc.}$  
So the excitation states in spin-orbital sector include both
the singlet $(0,0,0)$  and the multiplets of 15-fold $(1,0,1)$, of 20-fold
$(0,2,0)$, and of 45-fold $(2,1,0)$ or of 35-fold $(4,0,0)$ etc.
After evaluating the contributions of roots and two-strings to the 
highest weight vectors that characterize the irreducible representations
of SU(4), we can get the correct compositions of holes and two-strings
that create the possible excitations allowed by group 
theory. 

\subsection{The multiplets}

One $\lambda$ hole and one $\nu$ hole together create a 15-fold multiplet.
Let $\sigma_1^{(o)}(\lambda)
=\sigma^h(\lambda)=\delta(\lambda-\bar{\lambda})$,
$\tau_1^{(o)}(\nu)=\tau^h(\nu)=\delta(\nu-\bar{\nu})$, 
and the other inhomogeneous terms in eq.(\ref{eq:densityExcite}) be null.
Equation (\ref{eq:densityExcite}) is reduced to a closed form
by Fourier transform. 
The excitation energy is composed of two terms
$$\Delta E_{(15)} = \varepsilon_\sigma (\bar{\lambda})
            +\varepsilon_\tau(\bar{\nu}),$$
and each of them can be identified as a flavoron with energy
\begin{equation}
\varepsilon_f(\bar{x})=
  -\int_{-k_F}^{k_F}dk (2t\cos k+\Lambda)
     \rho^f_1 (k, \bar{x}),
\label{eq:flavoron}
\end{equation}
where ${\small f}=\sigma$, $\tau$, or $\omega$. 
The $\rho_1^\sigma(k,\bar{\lambda})$ is solved by   
\begin{eqnarray}
\rho_1^\sigma(k,\bar{\lambda})
 +\frac{\cos k/(4c)}{\sqrt{2}
   \cosh[\displaystyle\frac{\pi}{2c}(\sin k-\bar{\lambda})]-1}=
           \nonumber\\[3mm]
\frac{\cos k}{c}\int_{-k_F}^{k_F}R_{3/2}
    \left(\frac{\sin k-sin k'}{c}\right)
     \rho_1^\sigma(k',\bar{\lambda})dk',
\label{eq:lambdarho}
\end{eqnarray}
and the $\rho_1^\tau(k,\bar{\nu})$ satisfies the following
integral equation:
\begin{eqnarray}
\rho_1^\tau(k,\bar{\nu})
 +\frac{\cos k/(4c)}{\sqrt{2}
   \cosh[\displaystyle\frac{\pi}{2c}(\sin k-\bar{\nu})]+1}=
           \nonumber\\
\frac{\cos k}{c}\int_{-k_F}^{k_F}R_{3/2}
 \left(\frac{\sin k-sin k'}{c}\right)
  \rho_1^\tau(k',\bar{\nu})dk'.
\label{eq:nurho}
\end{eqnarray}

Two $\mu$ holes, 
$\omega^{(0)}(\mu)=\delta(\mu-\bar{\mu}_1)+\delta(\mu-\bar{\mu}_1)$ 
create a 20-fold multiplet with excitation energy
$$
\Delta E_{(20)}=
  \varepsilon_\omega (\bar{\mu}_1)
   +\varepsilon_\omega (\bar{\mu}_2),
$$
where $\varepsilon_\omega (x)$ is evaluated by the same
integral (\ref{eq:flavoron}), 
but $\rho_1^\omega(k,\bar{\mu})$ should solve
\begin{eqnarray}
\rho_1^\omega(k,\bar{\mu})
 +\frac{\cos k/(4c)}{
   \cosh[\displaystyle\frac{\pi}{2c}(\sin k-\bar{\mu})]}=
           \nonumber\\
\frac{\cos k}{c}\int_{-k_F}^{k_F}R_{3/2}
 \left(\frac{\sin k-sin k'}{c}\right)
  \rho_1^\omega(k',\bar{\mu})dk'.
\label{eq:murho}
\end{eqnarray}

The 45-fold multiplet is a three-hole state created by 
two $\lambda$ holes and one $\mu$ hole, i.e., 
$\sigma_1^{(o)}(\lambda)=\delta(\lambda-\bar{\lambda}_1)
 +\delta(\lambda-\bar{\lambda}_2)$ and
$\omega_1^{(o)}(\mu)=\delta(\mu-\bar{\mu})$
for which the excitation
energy is
$$ 
\Delta E_{(45)}=
 \varepsilon_\sigma(\bar{\lambda}_1)
 +\varepsilon_\sigma(\bar{\lambda}_2)
  +\varepsilon_\omega(\bar{\mu}).
$$

Four $\lambda$ holes create a 35-fold multiplet with excitation energy
$$
\Delta E_{(35)}=\sum_{j=1}^4
 \varepsilon_\sigma(\bar{\lambda}_j).
$$

In the above we have seen that there are three types of 
elementary excitation modes in the spin-orbital sector(let us
call the SU(4) flavor degree of freedom).
We refer these three elementary excitation modes the flavorons.
It is easy to know the contributions of the holes to the 
highest weight vectors, and to the spin and orbital.
Consequently,
the quadruplets $(1, 0, 0)$ or $(0, 0, 1)$ are
flavorons carrying both spin $1/2$ and orbital $1/2$
with energies $\epsilon_\sigma$ or $\epsilon_\tau$,
whereas the  hexaplet $(0, 1, 0)$ is a flavoron 
carrying either spin $1$ or orbital $1$ with energy 
$\epsilon_\omega$.
Clearly, spins and orbitals are mixed up in present isotropic 
on-site coupling. The spin orbital separation is expected to occur
for the anisotropic cases that can be caused by Hund's rule.
From eq.(\ref{eq:lambdarho}-\ref{eq:murho}) 
we find that the asymptotic behavior of 
all the three densities of roots vanish as the rapidities go
to infinity. Thus these elementary excitations are gapless,
i.e. $\epsilon_f(\pm\infty)=0$.

\subsection{The singlet}

By observing the contributions of two-strings to the highest
weight vectors, we find that the flavorons can compound to form a singlet.
In addition to placing to the $\lambda$-rapidity a hole at $\bar{\lambda}$
and to the $\nu$-rapidity a hole at $\bar{\nu}$, we take into account
of three two-strings respectively in those three rapidities,
$\lambda^{\pm}=\lambda_0\pm ic/2$, 
$\mu^{\pm}=\mu_0\pm ic/2$ and $\nu^{\pm}=\nu_0\pm ic/2$.
In this case, 
\begin{eqnarray}
M/L &=& \int_{-\infty}^{\infty}\sigma(\lambda)d\lambda+2,
    \nonumber\\
M'/L &=& \int_{-\infty}^{\infty}\omega(\mu)d\mu+2,
    \nonumber\\
M''/L &= &\int_{-\infty}^{\infty}\tau(\nu)d\nu+2,
  \nonumber
\end{eqnarray}
and the inhomogeneous terms of eq.(\ref{eq:densityExcite}) read
\begin{eqnarray}
\rho_1^{(o)}(k)
  &=&-\cos k K_1(\sin k-\lambda_0),
   \nonumber\\
\sigma_1^{(o)}(\lambda)
  &=&\delta(\lambda-\bar{\lambda})+K_{3/2}(\lambda-\lambda_0)
    \nonumber\\
     \,&\,& +K_{1/2}(\lambda_-\lambda_0)-K_1(\lambda-\lambda_0),
      \nonumber\\
\omega_1^{(o)}(\mu)
  &=&K_{3/2}(\mu-\mu_0)+K_{1/2}(\mu_-\mu_0)
   \nonumber\\
    \,&\,&-K_1(\mu-\lambda_0)-K_1(\mu-\nu_0),
     \nonumber\\
\tau_1^{(o)}(\nu)
 &=&\delta(\nu-\nu_0)+K_{3/2}(\nu-\nu_0)
  \nonumber \\
   \,&\,& +K_{1/2}(\nu-\nu_0)-K_1(\nu-\mu_0).
\label{eq:holeStrong}
\end{eqnarray}
Substituting them into eq.(\ref{eq:densityExcite}) and taking Fourier
transform, we find that the term containing $\nu_0$ in the fourth
equation cancels with the term containing $\nu_0$ in the third equation. 
After substituting
the result into the second equation, again the terms containing $\mu_0$
cancel each other. The substituting of the obtained expression of 
$\sigma_1(\lambda)$ into the first equation brings about an exact 
cancellation of the terms containing $\lambda_0$. As a result,
the excitation energy is obtained 
$$\Delta E_{(1)}= \varepsilon_\sigma(\bar{\lambda})
 +\varepsilon_\tau(\bar{\nu}),$$
where $\varepsilon_\sigma$ and $\varepsilon_\tau$ 
are evaluated by eq.(\ref{eq:flavoron})
in which the $\rho_1^\sigma$ and $\rho_1^\tau$
satisfy the eq.(\ref{eq:lambdarho}) and  eq.(\ref{eq:nurho}), respectively. 
Clearly, the excitations of 15-fold multiplet and the singlet
are degenerate in energy.

\section{Charge excitations}\label{sec:charge}

\subsection{The holon-antiholon excitation}

Let us consider the case of less than quarter-filling ($N<L$).
we are allowed to add one ``particle'' outside the charge Fermi sea,
$k_p\notin[-k_F, k_F]$ but leaving a hole inside the charge Fermi sea,
$\bar{k}\in[-k_F, k_F]$. The calculation of the energy is 
required  to start from 
\[
E=-2t\cos k_p -2tL\int_{-k_F}^{k_F}\cos k \rho(k)dk,
\]
where the integration limit $k_F$ is required to fulfill 
$\int_{-k_F}^{k_F}\rho(k)dk=(N-1)/L$.

By introducing $\rho(k)=\rho_0(k)+\rho_1(k)/L$ etc., 
the excitation energy $\Delta E=E-E_0$ is composed of two terms
\begin{equation}
\Delta E(\bar{k},k_p)=\varepsilon_h(\bar{k})
     +\bar{\varepsilon}_h(k_p).
\label{eq:h-pExcite}
\end{equation}
Here we introduced the holon energy
\begin{eqnarray}
\varepsilon_h(x)= 2t\cos x 
   -\int_{-k_F}^{k_F}(2t\cos k+\Lambda)\rho^c_1(k,x)dk,
\label{eq:holon}
\end{eqnarray}
and antiholon (``particle'' state) energy 
$\bar{\varepsilon}_h(x)=-\varepsilon_h(x)$.
In eq.(\ref{eq:holon}), the $\rho_1^c(k,x)$ should be solved 
from the following equation
\begin{eqnarray}
\rho_1^c(k,x)
 +\frac{\cos k}{c}R_{3/2}
  \left(\frac{\sin k-\sin x}{c}\right)= 
   \nonumber\\
\frac{\cos k}{c}\int_{-k_F}^{k_F}R_{3/2}
 \left(\frac{\sin k-sin k'}{c}\right)
  \rho_1^c(k',x)dk'.
\label{eq:krho}
\end{eqnarray}
This equation is derived from (\ref{eq:densityExcite}) by taking
\begin{eqnarray}
\rho_1^{(o)}(k)&=&\delta(k-\bar{k}), \nonumber\\
\sigma_1^{(o)}(\lambda)&=&-K_{1/2}(\lambda-\sin k_p), \nonumber\\
\omega_1^{(o)}(\mu)&=&\tau_1^{(o)}(\nu)=0, \nonumber 
\end{eqnarray}
which comes from the ``particle'' and the hole.
Now we find that this kind of excitation consists of a holon carrying
energy $\varepsilon(\bar{k})$ and an antiholon carrying energy 
$\bar{\varepsilon}_h(k_p)$. Obviously, eq.(\ref{eq:h-pExcite}) 
vanishes when $\bar{k}\rightarrow k_F$ and $k_p\rightarrow k_F$,
and the holon-antiholon excitation is gapless.

\subsection{The holon-holon excitation}

In order to discuss states with double occupancy, we need to 
consider solutions
containing complex $k$ pairs. 
Suppose a configuration $\{h_j\}$ results in a 
complex pairs (two-strings), $k^\pm=\kappa\mp i\chi$ and two holes inside the
Fermi-sea, $\bar{k}_1$, $\bar{k}_2 \in [-k_F, k_F]$. After a careful
analyses of the Bethe-ansatz equation, one finds the string position is
restricted to lie around a particular $\lambda$ solution that we denoted
by $\lambda_0$, i.e., it must satisfy 
\[
\sin(\kappa\pm i\chi)=\lambda_0\mp ic/2+O(e^{-\eta L}).
\]
An exact deletion holds in the second equation of (\ref{eq:BAE}) for
$\lambda_0=(\sin\bar{k}_1+\sin\bar{k}_2)/2$. After rewriting the 
Bethe-ansatz equation by separating the factors of the complex $k$ pairs,
we take the thermodynamics limit as before. In order 
to get the excitation energy
we need to solve (\ref{eq:densityExcite}) with
\begin{eqnarray}
\rho_1^{(o)}(k)&=&\delta(k-\bar{k}_1)+\delta(k-\bar{k}_2)\nonumber\\
   \,&\,& +\cos k \, K_{-1/2}(\sin k-\lambda_0),\nonumber\\
\sigma_1^{(o)}(\lambda)&=&\tau_1^{(o)}(\nu)=0, \nonumber\\
\omega_1^{(o)}(\mu)&=&K_{-1/2}(\mu-\lambda_0).\nonumber
\end{eqnarray} 
After careful calculation, we obtain the excitation energy
\[
\Delta E(\bar{k}_1, \bar{k}_2)=\varepsilon_h(\bar{k}_1)
    +\varepsilon_h(\bar{k}_1)+\Delta(U, k_F),
\]
where $\varepsilon_h$ is the holon energy given by the same equation 
(\ref{eq:holon}), and $\Delta(U,k_F)$ is given by
\begin{eqnarray}
\Delta(U, k_F)=U+2t\int_{-\pi}^{\pi}\cos^2k K_1(\sin k-\lambda_0)dk
  \nonumber\\
-\int_{-k_F}^{k_F}(2t\cos k+\Lambda)\rho_1^{stri}(k,\lambda_0)dk,
  \nonumber
\end{eqnarray}
with 
\begin{eqnarray}
\,&\,&\hspace{2mm}\rho_1^{stri}(k,\lambda_0)=
 \frac{\cos k}{c}R_1\left(\frac{\sin k-\lambda_0}{c}\right) 
  \nonumber\\
&+&\frac{\cos k}{c}\int_{-k_F}^{k_F}R_{3/2}
  \left(\frac{\sin k-sin k'}{c}\right)
   \rho_1^{stri}(k',\lambda_0)dk'
    \nonumber\\
&+&\frac{\cos k}{c}K_{1/2}(\sin k-\lambda_0).  
\label{eq:stringrho}
\end{eqnarray}
Clearly the holon-holon excitation always has a gap 
$\Delta_g=2\varepsilon_h(k_F)+\Delta(U,k_F)$, which exists 
at any filling. However, the gapless modes of holon-antiholon 
are available to carry charges for away from quarter-filling.
It is easy to shown by calculating (\ref{eq:HWintegral})
that both holon-antiholon and holon-holon excitations 
are SU(4) singlet, consequently, they carry neither spins nor orbitals.

\section{Discussions}\label{sec:conclusion}

In the above we have presented an extensive discussion on one-dimensional
Hubbard-like model with SU(4) symmetry,
where the sites are restricted to 
be occupied by at most two electrons.. 
The model was proposed to describe electrons with
two-fold orbital degeneracy. The symmetries and some general features 
were given previously \cite{LiE99}. 
We focused on the one dimensional case in this paper and studied 
the ground state and excitations by means of an exact solution.
The excitation energies of the excited states are just sums of some 
particular terms related to quasiparticles.
It provides an explicitly interpretation of the 
separation of charge excitations
and spin-orbital excitations. Among the charge excitations, there are 
gapless holon-antiholon excitations and holon-holon excitations with gap. 
Both excitations carry neither spins nor orbitals. 
They are completely decoupled from the spin-orbital degree of freedom. 
The holons and antiholons move throughout the crystal at less than  
quarter-filling. Various excitations in spin-orbital sector consist 
of three basic modes which are created by the holes in the three
rapidities for the spin-orbital double. 
That means there are three kinds of quasiparticles 
that carry spins and orbitals,
i.e., two quadruplets transforming according to
the fundamental or conjugate representation, and one
hexaplet forming the six-dimensional representation.
These elementary excitations in 
spin-orbital sector are gapless.

As the on-site coupling in our model
is isotropic for spin-orbital labels, there is
no separation between spin and orbital. A complete separation 
between the spin waves and orbital waves is expected to occur after taking
account of the contributions of Hund's rule. This needs to introduce 
anisotropic on-site coupling in the spin and orbital configuration.

For finite $N$ and $L$ we plotted the excitation spectra by solving the
Bethe-ansatz equation numerically. The variation of the quantum number
for excited states from that for the ground state, and their changes from 
integer to half-integers (or vice versa)  
were shown in each cases. It provides a concrete
interpretation about the collective excitations for the orbital degenerate
electronic systems. 
The overall structure of the spectra for spin-orbital excitations
changed greatly  with respect to the changes of the correlation strength.
The lowest excitation energy and the whole pattern are raised 
when the correlation strength decreased. 
However, the ``particle''-hole excitation spectrum
does not change much from the strong to the weak 
correlation strengths.

In the quarter-filled band for strong repulsive on-site coupling,
there will be no doubly occupied sites. In this case the total wave
function will be separated into a product of Slater determinant of 
$N$ ``spinless'' fermions and part of SU(4) 
Heisenberg magnets. The direct results from the Bethe-ansatz equation by
taking strong coupling limit agree with it exactly.

It is worthwhile to mention that the model studied here is not a direct
SU(4) generalization of the Hubbard model, since a projection onto
the subspace of states having at most two electrons at each site
was made to render it solvable through the 
Bethe-ansatz. The two models are therefore
not expected to share the same physical features. Considering the
self-conjugate representation on a bipartite lattice in the strong repulsive
coupling limit, Ref.\cite{Afflecketal} clarified the system is dimerized 
with doubly degenerate singlet ground state and indicated
the excitations are massive symmetry and antisymmetric kinks.
In our present model, however, the local states on each site carry
out the fundamental representation of SU(4). 

YQL acknowledges the supports of AvH Stiftung and
interesting discussions with H Frahm. 
This work is supported by NSFC-19975040 and EYF of China Education Ministry.

\end{multicols}


\newpage

\newpage

\newpage

\end{document}